\documentclass[twocolumn]{aastex701}

\usepackage{graphicx}
\usepackage{amsmath,amssymb}
\usepackage{physics}
\usepackage{booktabs}
\usepackage{siunitx}
\usepackage{enumitem}
\usepackage{multirow}
\usepackage{tikz}
\usepackage{ragged2e}
\usepackage{aas_macros}

\graphicspath{{./}{fig/}}

\newcommand{\nishiura}[1]{\textcolor{black}{#1}}
\newcommand{\Rei}[1]{}

\shorttitle{Bell Instability and CR Acceleration in AGN UFO Shocks}
\shortauthors{Nishiura \& Inoue}

\begin{document}

\title{Bell Instability and Cosmic-Ray Acceleration in Active Galactic Nuclei Ultrafast Outflow Shocks}

\correspondingauthor{Rei Nishiura}

\author[orcid=0009-0003-8209-5030,gname=Rei,sname=Nishiura]{Rei Nishiura}
\affiliation{Department of Physics, Kyoto University, Kyoto 606-8502, Japan}
\email[show]{nishiura@tap.scphys.kyoto-u.ac.jp}   

\author[orcid=0000-0002-7935-8771,gname=Tsuyoshi,sname=Inoue]{Tsuyoshi Inoue}
\affiliation{Department of Physics, Konan University, Okamoto 8-9-1, Higashinada-ku, Kobe 658-8501, Japan}
\email{tsuyoshi.inoue@konan-u.ac.jp}


\begin{abstract}
We investigate magnetic-field amplification driven by the nonresonant hybrid (NRH or Bell) instability and its impact on cosmic-ray (CR) acceleration at reverse shocks of ultrafast outflows (UFOs) from active galactic nuclei (AGN). Previous kinetic studies by particle-in-cell simulations have demonstrated that when maximum CR energy is near the injection scale, NRH instability efficiently amplifies magnetic field  up to the saturation level. However, the efficiency of NRH instability goes down as \nishiura{the }maximum energy increase\nishiura{s} since CR current is carried by escaping CRs near the maximum energy. We employ a one-dimensional MHD--CR framework solving telegraph-type diffusion--convection equations to trace the coupled evolution of CRs, magnetic fields, and shock dynamics under realistic parameters. We find a distinct transition with magnetic field strength: for weak background fields ($B_{0}\!\lesssim\!10^{-4}\,\mathrm{G}$), NRH instability efficiently amplifies upstream turbulence, driving a self-regulated state where $E_{\max}$ becomes independent of initial strength of magnetic turbulence. In contrast, for stronger background fields ($B_{0}\!\gtrsim\!10^{-3}\,\mathrm{G}$), the escaping CR current is too weak to drive NRH instability, and magnetic turbulence further decays through parametric instabilities, potentially reducing the acceleration efficiency. We give the physical interpretation for the transition and discuss conditions for PeV--EeV acceleration at UFO reverse shocks.
\end{abstract}




\section{INTRODUCTION}
Cosmic rays (CRs) are high-energy charged particles propagating through interstellar and intergalactic space, spanning energies from $10^{9}\,\mathrm{eV}$ to $10^{20}\,\mathrm{eV}$. Understanding their origin is not only a fundamental problem in high-energy astrophysics but also essential for galaxy evolution. Despite extensive studies, the origin of CRs across different energy ranges remains unresolved. Supernova remnants (SNRs) are regarded as the most promising candidates up to the knee energy (a few $\times10^{15}\,\mathrm{eV}$) \citep{1934PhRv...46...76B,1978MNRAS.182..147B,1978ApJ...221L..29B}, whereas the sources and acceleration mechanisms at and above the knee energy are largely unknown.

Active galactic nuclei (AGN) are among the leading candidates for the origin of high-energy CRs across and beyond the knee energy. Proposed acceleration sites in AGN include relativistic jets \citep{2004NewAR..48..381A,2014PhRvD..90b3007M,2015SSRv..191..519S,2018ApJ...863L..10A,2018ApJ...865..124M}, coronae \citep{2020ApJ...891L..33I,2020PhRvL.125a1101M,2021ApJ...922...45K,2022ApJ...939...43E,2022ApJ...941L..17M}, accretion flows \citep{2019PhRvD.100h3014K,2021A&A...649A..87G}, and outflows \citep{2016A&A...596A..68L,2017PhRvD..95f3007W,2018ApJ...858....9L,2022arXiv220702097I,2023MNRAS.526..181P}. In addition, AGN are considered promising multi-messenger sources because the acceleration of CRs is expected to produce high-energy neutrinos and gamma rays \citep{2017PhRvD..95f3007W,2018ApJ...858....9L,2020PhRvL.124e1103A,2021ApJ...921..144A,2022Sci...378..538I,2022arXiv220702097I,2025JCAP...07..013P,2025ApJ...980..131S}.

Among the AGN environments, ultrafast outflows (UFOs) have recently attracted growing attention. UFOs are mildly relativistic winds, launched at tens of percent of the speed of light, and identified through blueshifted Fe K-shell absorption features (Fe XXV He$\alpha$ at 6.7 keV and Fe XXVI Ly$\alpha$ at 7.0 keV) or P-Cygni--like profiles. They are observed in several tens of percent of AGN \citep{2010A&A...521A..57T,2015MNRAS.451.4169G}, suggesting that they are more common than powerful jets. Observationally, the detection of GeV gamma-rays in some AGN---often difficult to explain by AGN coronae alone---provides further motivation to consider UFOs as promising acceleration sites \citep{2024ApJ...965...68M,2025ApJ...980..131S}. \nishiura{Recent Fermi-LAT stacking analyses of AGNs hosting UFOs have also found evidence for average GeV gamma-ray emission that is likely associated with the interaction between UFOs and the ambient medium \citep{2021ApJ...921..144A}.}

UFOs colliding with ambient gas are expected to form shock structures. These shocks have been proposed as promising sites for diffusive shock acceleration (DSA) \citep{1978MNRAS.182..147B,1978ApJ...221L..29B,1983RPPh...46..973D,1987PhR...154....1B,2016RPPh...79d6901M}, in which CRs repeatedly cross the shock front via pitch-angle scattering off magnetic turbulence, gaining energy efficiently. Recent studies indicate that protons may be accelerated up to $10^{18}\,\mathrm{eV}$ and heavy nuclei up to $10^{20}\,\mathrm{eV}$ at reverse shocks of the UFOs \citep{2023MNRAS.526..181P,2025MNRAS.539.2435E}.

The maximum energy achievable via DSA depends primarily on magnetic-field strength, turbulence amplitude, and shock velocity (Eq.~\eqref{eq:order_estimation_maximum_energy_CR_Bell}). Owing to their mildly relativistic high velocities, UFO shocks are expected to accelerate CRs efficiently. However, observational constraints on the magnetic-field strength remain largely uncertain. Many theoretical studies therefore introduce a phenomenological parameter, $\epsilon_B$, to characterize the fraction of shock energy in magnetic fields. Moreover, \nishiura{Bohm-like diffusion in the limit of strong turbulence with $\delta B/B_0\gtrsim1$} are often assumed without explicitly treating the physics of magnetic field amplification, and saturation. Such simplifications potentially introduce systematic uncertainties in estimates of $E_{\max}$ and, consequently, predictions of neutrino and gamma-ray emission.

The nonresonant hybrid instability (NRH instability, also known as the Bell instability) offers a self-consistent mechanism for determining key magnetic-field parameters such as $\varepsilon_B$ and $\delta B/B_0$ in the shock vicinity \citep{2004MNRAS.353..550B}. This instability is driven by the CR current escaping upstream of the shock, which induces a return current in the background plasma and amplifies circularly polarized Alfvén waves. Magnetic-field amplification regions have been observed in SNRs \citep{2003ApJ...584..758V,2003ApJ...589..827B,2005ApJ...621..793B}, and these are often interpreted as possible evidence of NRH instability.

NRH instability has been investigated both analytically and numerically. Analytically, \citet{2004MNRAS.353..550B} derived the linear growth theory. Numerically, a wide range of methods has been applied, including full particle-in-cell (PIC) simulations \citep{2015PhRvL.114h5003P}, kinetic hybrid approaches \citep{2014ApJ...783...91C,2014ApJ...794...46C}, MHD-PIC methods \citep{2012ApJ...759...73N}, and hybrid MHD–Vlasov-Fokker-Planck (VFP) models \citep{2013MNRAS.431..415B,2013MNRAS.430.2873R}. While PIC simulations capture kinetic effects from shock formation to nonthermal injection, they are computationally prohibitive for tracing particle acceleration orders of magnitude larger than the injection. Hybrid MHD–VFP models reduce computational costs by expanding the CR Boltzmann equation into multipoles, but these studies in most case\nishiura{s} fixed the CR current, preventing a fully self-consistent treatment of particle acceleration and field amplification.

\citet{2019ApJ...872...46I,2021ApJ...922....7I,2024ApJ...965..113I} developed a numerical code capable of simultaneously evolving the background plasma and CR current in one-dimensional setups. Their results demonstrated that finite spatial extent of the upstream CR current supresses the growth of NRH instability compared to previous theoretical models that assume infinite spatial extent of the CR current. Nevertheless, under favorable conditions, NRH instability can amplify magnetic fields sufficiently to enable $10^{15}~\mathrm{eV}$ acceleration in very early stage of young SNRs.

In this work, we employ the numerical code of \citet{2019ApJ...872...46I} to investigate the growth and saturation of NRH instability in the context of UFOs. By systematically varying control parameters such as background magnetic field $B_0$, injection rate $\eta$, and initial amplitude of the magnetic fluctuation $\xi_{B,\mathrm{ini}}$ defined by Eq. \eqref{eq:definition_magnetic_fluctuation_strength}, we assess whether the maximum CR energy $E_{\max}$ in UFO shocks can be determined self-consistently, without resorting to phenomenological parameters.

The structure of this paper is as follows. In Sec.~\ref{sec:setup_UFO}, we describe the governing MHD–CR Boltzmann equations, diffusion coefficient, and the adopted boundary and initial conditions, including $B_0$, $\xi_{B,\mathrm{ini}}$, $\eta$, and $\mathrm{p}\gamma$ cooling. In Sec.~\ref{sec:result_UFO}, we present the main results: (i) simulations without the NRH term reproduce \nishiura{test-particle DSA}, (ii) in weak-field regimes ($B_0=10^{-5}$–$10^{-4}\,\mathrm{G}$), NRH instability amplifies the magnetic field, driving $E_{\max}$ and $\varepsilon_B$ to converge regardless of $\xi_{B,\mathrm{ini}}$, (iii) in strong-field regimes ($B_0\gtrsim10^{-3}\,\mathrm{G}$), NRH instability is ineffective and $E_{\max}$ depends on the initial conditions, (iv) variations in interstellar medium (ISM) density affect acceleration efficiency, and (v) $\mathrm{p}\gamma$ cooling can limit the maximum energy.

\section{Basic Equations and Simulation Setup}
\label{sec:setup_UFO}

To investigate particle acceleration in UFOs, we adopt the following physical assumptions. More detailed initial conditions are presented in Sec.~\ref{subsec:initial_condition_UFO}.

\begin{enumerate}[label=(\roman*)]
 \item When a UFO collides with the ISM, it generates a shock structure composed of a reverse shock, contact discontinuity, and forward shock. In this study, we focus on the local region around the reverse shock and analyze particle acceleration there (see Fig.~\ref{fig:shock_structure}). \nishiura{This choice is motivated by previous studies that proposed the UFO reverse shock as a promising site for acceleration up to EeV energies \citep{2023MNRAS.526..181P,2025MNRAS.539.2435E}. Moreover, in the upstream rest frame, the forward shock decelerates with time, whereas the reverse shock remains comparable to the wind velocity, so we may expect more favorable for efficient particle acceleration.}
 \item \nishiura{We define the upstream of the reverse shock as the wind region and the downstream as the shocked wind layer between the reverse shock and the contact discontinuity. The NRH instability is mainly excited in this upstream (wind) region (see Fig.~\ref{fig:magnetic_fluctuation_from_shock_10-5}).}
 \item CRs are assumed to be pure protons.
 \item The wind is assumed to be steady and uniform\footnote{Variability of UFOs has been observed on timescales from months down to days \citep{2008MNRAS.385L.108R,2009A&A...504..401C,2009MNRAS.397..249P}. Moreover, both theoretical and observational studies suggest that UFOs are clumpy and inhomogeneous \citep{2013PASJ...65...88T,2025Natur.641.1132X}. These factors may affect particle acceleration and will be considered in future work.}.
 \item The background magnetic field is assumed to be
uniform in the $x$ direction,
\begin{equation}
B_x \equiv B_0 = \mathrm{const},
\label{eq:background_magnetic_field_UFO}
\end{equation}
\nishiura{and parallel to the shock normal over a coherence length $\ell_{\mathrm{coh}}$.} In addition, broadband, circular-polarized Alfv\'en wave turbulence is superposed on the $y$ and $z$ components of magnetic field (see Sec.~\ref{subsec:initial_magnetic_field} for detail). \nishiura{Although Parker-type wind solutions imply that the large scale magnetic field becomes toroidally dominated at large radii, UFO outflows are expected to be intermittent and turbulent rather than laminar. In such a turbulent dynamo, the magnetic field is amplified on the hydrodynamic eddy scale and naturally forms local patches with a coherent mean field. The NRH instability can grow when the coherence length of $B_x$ exceeds the wavelength of the NRH mode with maximum growth, as given by Eq.~\eqref{eq:maximum_growth_wavelength_Bell}, that is, when $\ell_{\mathrm{coh}} > \lambda_{\mathrm{NRH}}$. Even if the large scale geometry of the UFO reverse shock is quasi-perpendicular, such locally quasi-parallel patches with a coherent background magnetic field are expected to exist in a turbulent dynamo. In our setup, the initial condition $\delta B/B_0 < 1$ is imposed at the NRH maximum growth scale, so that the NRH mode effectively samples a nearly uniform background magnetic field with $\ell_{\mathrm{coh}}> \lambda_{\mathrm{NRH}}$, while the field can still be strongly tangled on larger hydrodynamic scales.
}
 \item The spatial dependence of physical quantities is restricted to one dimension along the $x$-axis, and curvature effects are neglected. On the propagation timescale of UFOs across $\sim\mathrm{pc}$ scales, curvature effects on both fluid dynamics and CR transport are negligible.
\end{enumerate}

\subsection{Basic Equations}
In this study, we solve the coupled system of magnetohydrodynamic (MHD) equations for the fluid component and telegraph-type diffusion--convection equations for the CR distribution function \citep{2013MNRAS.431..415B,2019ApJ...872...46I,2021ApJ...922....7I,2024ApJ...965..113I}. The MHD equations are expressed as follows. The continuity equation is given by
\begin{equation}
\frac{\partial \rho}{\partial t} + \frac{\partial (\rho v_x)}{\partial x} = 0.
\label{eq:continuity_equation_Bell}
\end{equation}
The momentum equations are
\begin{equation}
\frac{\partial (\rho v_x)}{\partial t} + \frac{\partial}{\partial x} \left(\rho v_x^2 + P + \frac{B_y^2 + B_z^2}{8 \pi}\right) = 0,
\label{eq:EOM_Bell_x}
\end{equation}
\begin{equation}
\frac{\partial (\rho v_y)}{\partial t} + \frac{\partial}{\partial x} \left(\rho v_x v_y - \frac{B_x B_y}{4 \pi}\right) = -\frac{1}{c} j_x^{(\mathrm{ret})} B_z,
\label{eq:EOM_Bell_y}
\end{equation}
\begin{equation}
\frac{\partial (\rho v_z)}{\partial t} + \frac{\partial}{\partial x} \left(\rho v_z v_x - \frac{B_z B_x}{4 \pi}\right) = \frac{1}{c} j_x^{(\mathrm{ret})} B_y.
\label{eq:EOM_Bell_z}
\end{equation}
The energy equation is represented by
\begin{equation}
\frac{\partial \epsilon}{\partial t} + \frac{\partial}{\partial x} \left\{ v_x \left( \epsilon + P + \frac{B_y^2 + B_z^2}{8 \pi} \right) - B_x \frac{\bm{B} \cdot \bm{v}}{4 \pi} \right\} = 0.
\label{eq:energy_equation_Bell}
\end{equation}
The induction equations for $B_y$ and $B_z$ are
\begin{equation}
\frac{\partial B_y}{\partial t} = \frac{\partial}{\partial x} \left( B_x v_y - B_y v_x \right),
\label{eq:induction_equation_Bell_y}
\end{equation}
\begin{equation}
\frac{\partial B_z}{\partial t} = \frac{\partial}{\partial x} \left( B_x v_z - B_z v_x \right).
\label{eq:induction_equation_Bell_z}
\end{equation}
Here, the total energy density is defined as
\begin{equation}
\epsilon \equiv \frac{P}{\gamma-1}+\frac{1}{2} \rho v^2+\frac{B_y^2+B_z^2}{8 \pi},
\label{eq:definition_energy_density}
\end{equation}
which includes internal, kinetic, and magnetic energies, where $\gamma$ is the adiabatic index. The return current $j_x^{(\mathrm{ret})}=-j_x^{(\mathrm{CR})}$ maintains charge neutrality against the CR current $j_x^{(\mathrm{CR})}$. It is expressed in terms of the anisotropic part of the distribution function $f_1$, as defined later in Eq.~\eqref{eq:return_current_Bell}.

\begin{figure}
\RaggedRight
\includegraphics[width=\columnwidth]{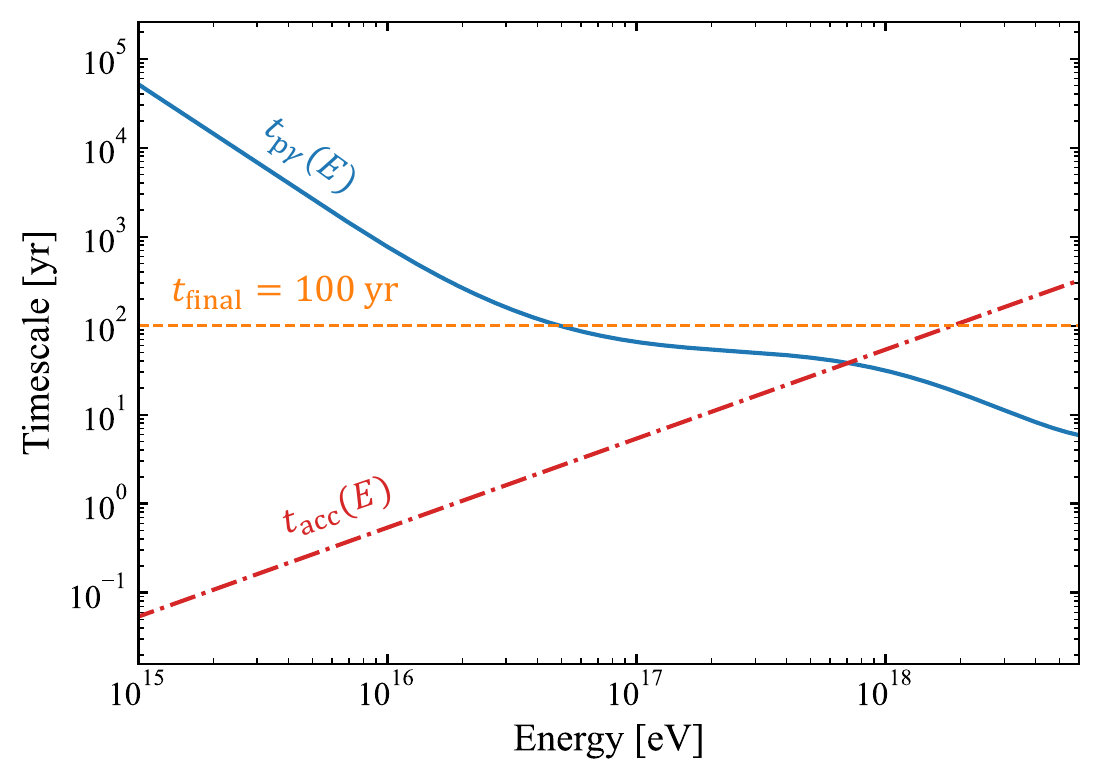}
\caption{\justifying Comparison of $\mathrm{p}\gamma$ cooling and acceleration timescales as a function of proton energy. The blue solid line shows the $\mathrm{p}\gamma$ cooling timescale, the orange dashed line indicates the simulation runtime (the propagation timescale of the reverse shock across pc scales), and the red dot–dashed line represents the acceleration timescale described in Sec.~\ref{subsec:pgamma_cooling_UFO}, calculated from Eq.~\eqref{eq:acceleration_time_UFO} with $\xi_{B}=0.1$, $B_0=10^{-2}~\mathrm{G}$, and $v_{\text{sh}}=5.0\times10^9~\mathrm{cm~s^{-1}}$ (see Tab.~\ref{tab:Model_Parameters_UFO} for the fiducial model).}
\label{fig:pgamma_cooling_timescale}
\end{figure}

The CR distribution function is expanded with respect to the pitch-angle cosine relative to the background magnetic field $\bm{B}_0$ as follows,
\begin{equation}
f(x, \boldsymbol{p}) = f_0(x, p) + \frac{p_x}{p} f_1(x, p),
\label{eq:definition_of_CR_distribution_Bell}
\end{equation}
where $f_0(x,p)$ denotes the isotropic part and $f_1(x,p)$ the anisotropic part. The CR current can then be written as
\begin{equation}
j_x^{(\mathrm{CR})}=-j_x^{(\mathrm{ret})}=c e \int_{p_{\min }}^{p_{\max }} f_1(x, p) \frac{4 \pi}{3} p^2 \mathrm{~d} p.
\label{eq:return_current_Bell}
\end{equation}
By defining
\begin{equation}
F_0 \equiv f_0 p^3, \quad F_1 \equiv f_1 p^3,
\end{equation}
the evolution equations for $F_0$ and $F_1$ can be expressed as
\begin{equation}
\begin{aligned}
\frac{\partial F_0}{\partial t} 
&+ \frac{\partial\left(v_x F_0\right)}{\partial x} 
- \frac{1}{3} \frac{\partial v_x}{\partial x} 
  \frac{\partial F_0}{\partial \ln p}  \\
&= -\frac{c}{3} \frac{\partial F_1}{\partial x} 
+ Q_{\mathrm{inj}} p^3 
- t_{\mathrm{p}\gamma}^{-1}(p)F_0,
\end{aligned}
\label{eq:CR_evolution_equation1}
\end{equation}
\begin{equation}
\frac{\partial F_1}{\partial t} + \frac{\partial\left(v_x F_1\right)}{\partial x} = -c \frac{\partial F_0}{\partial x} - \frac{c^2}{3 D_{\|}(p, B)} F_1.
\label{eq:CR_evolution_equation2}
\end{equation}
Here, $Q_{\mathrm{inj}}$ is the CR injection rate, $t_{\mathrm{p}\gamma}(p)$ is the timescale of $\mathrm{p}\gamma$ interactions near the AGN source (see Fig.~\ref{fig:pgamma_cooling_timescale}), and $D_{\|}(p,B)$ is the diffusion coefficient. In this work, we incorporate the $\mathrm{p}\gamma$ energy-loss timescale derived by \citet{2023MNRAS.526..181P} into the numerical framework of \citet{2019ApJ...872...46I}. Note that Eqs. \eqref{eq:CR_evolution_equation1} and \eqref{eq:CR_evolution_equation2} are reduced to the diffusion convection equation in the limit $c\rightarrow \infty$ (see, \citet{2021ApJ...922....7I} for numerical tests).

The injection rate $Q_{\mathrm{inj}}$ is modeled as follows. A fraction $\eta$ of upstream fluid particles with momentum $p_{\text{inj}}$ are assumed to be injected into the acceleration process \citep{10.1111/j.1365-2966.2005.09227.x}:
\begin{equation}
Q_{\mathrm{inj}}(x, p)=\frac{\eta n_{\mathrm{wind}} v_{\mathrm{sh}} }{4 \pi p_{\text{inj} }^2} \delta\left(p-p_{\text{inj} }\right) \delta\left(x-x_{\mathrm{sh}}\right),
\end{equation}
where $n_{\mathrm{wind}}$ is the upstream wind density, $v_{\mathrm{sh}}$ is the shock velocity in the upstream rest frame, and $x_{\mathrm{sh}}$ is the shock position. The relation between $\eta$ and $p_{\text{inj}}$ is given by
\begin{equation}
\eta \equiv \frac{\int_{p_{\text {inj }}}^{\infty} \exp \left(-\frac{p^2}{2 m_{\mathrm{g}} k_{\mathrm{B}} T_{\text {sh }}}\right) \mathrm{d} p}{\int_0^{\infty} \exp \left(-\frac{p^2}{2 m_{\mathrm{g}} k_{\mathrm{B}} T_{\text {sh }}}\right) \mathrm{d} p},
\end{equation}
with $m_{\mathrm{g}}$ the mean gas mass and $T_{\text{sh}}$ the downstream temperature. For strong shocks with $\gamma=5/3$, the Rankine–Hugoniot relation gives $T_{\text{sh}}=3m_{\text{g}}v_{\text{sh}}/(16k_{\text{B}})$. We adopt $m_{\text{g}}=1.27m_{\text{p}}$, consistent with the solar composition inferred from Cygnus A emission-line ratios \citep[][Tab.~13.4]{2006agna.book.....O}. For numerical reasons, the momentum range of CR is restricted to $p_{\min }>p_{\mathrm{inj}}$ up to $p_{\max}$. Between $p_{\mathrm{inj}}$ and $p_{\min}$, we assume the standard DSA spectrum $f_0(x_{\text{sh}})\propto p^{-4}$ \citep{2019ApJ...872...46I,2021ApJ...922....7I,2024ApJ...965..113I}. Accordingly, the injection rate used in simulations is rewritten as
\begin{equation}
Q_{\mathrm{inj}}(x, p)=\frac{\eta n_{\mathrm{wind}} v_{\mathrm{sh}} p_{\mathrm{inj}}}{4 \pi p_{\min }^3}\frac{1}{\Delta p_{\text{min}}}\frac{1}{\Delta x},
\label{eq:injection_for_simulation}
\end{equation}
where the delta functions are replaced by grid intervals.

The diffusion coefficient is expressed as
\begin{equation}
D_{\|}(p, B) = \frac{4}{3 \pi} \frac{\max \left(B_0^2, \delta B^2\right)}{\delta B^2} \frac{c E_{\mathrm{CR}}}{e \max \left(B_0, \delta B\right)},
\label{eq:diffusion_coefficient_pitch_angle}
\end{equation}
where $\delta B^2 = B_y^2 + B_z^2$ and $E_{\mathrm{CR}}$ is the CR energy. For $\delta B < B_0$, Eq.~\eqref{eq:diffusion_coefficient_pitch_angle} reduces to the pitch-angle scattering coefficient, while for $\delta B \geq B_0$ it corresponds to the \nishiura{Bohm-like limit}.

The validity of Eq.~\eqref{eq:diffusion_coefficient_pitch_angle} is supported by two arguments. First, kinetic hybrid simulations \citep{2014ApJ...794...47C} and test-particle calculations \citep{2016APh....73....1R} have shown that NRH-amplified fields drive diffusion close to the \nishiura{Bohm-like limit}. Second, NRH instability alone amplifies only small-scale magnetic fluctuations, with wavelengths much shorter than the gyroradius at $E_{\max}$. Filamentation instability can transfer this energy to larger scales, amplifying turbulence up to the gyroradius scale \citep{2012MNRAS.419.2433R,2013ApJ...765L..20C}.

\subsection{Initial Conditions}
\label{subsec:initial_condition_UFO}
\begin{table*}[htbp]
  \centering
  \renewcommand{\arraystretch}{1.5}
  \sisetup{table-number-alignment=left}
  \caption{Simulation model parameters and categories. 
  The fiducial model explores a wide parameter range, 
  while the other models test the effects of ISM density 
  and $\mathrm{p}\gamma$ cooling.}
  \label{tab:Model_Parameters_UFO}
  \begin{tabular}{lcccccl}
    \toprule
    Model ID & {$n_{\text{ISM}}$ [$\mathrm{cm}^{-3}$]} & {$B_0$ [G]} & 
    {$\dot{M}$ [$M_{\odot}/\mathrm{yr}$]} & {$\eta$ (injection rate)} & {$\xi_{B,\mathrm{ini}}$} \\
    \midrule
    Fiducial & $10^2$ & $10^{-5}/10^{-4}/10^{-3}/10^{-2}$ & $0.1$ & $10^{-4}/10^{-3}$ & $0.1/0.01$ \\
    ISMlow   & $10^0$ & $10^{-5}$                         & $0.1$ & $10^{-4}$        & $0.1$ \\
    ISMhigh  & $10^4$ & $10^{-5}$                         & $0.1$ & $10^{-4}$        & $0.1$ \\
    p$\gamma$ & $10^2$ & $10^{-4}/10^{-3}/10^{-2}$        & $0.1$ & $10^{-4}$        & $0.1$ \\
    \bottomrule
  \end{tabular}
\end{table*}
\begin{figure*}
\centering
\includegraphics[width=\textwidth]{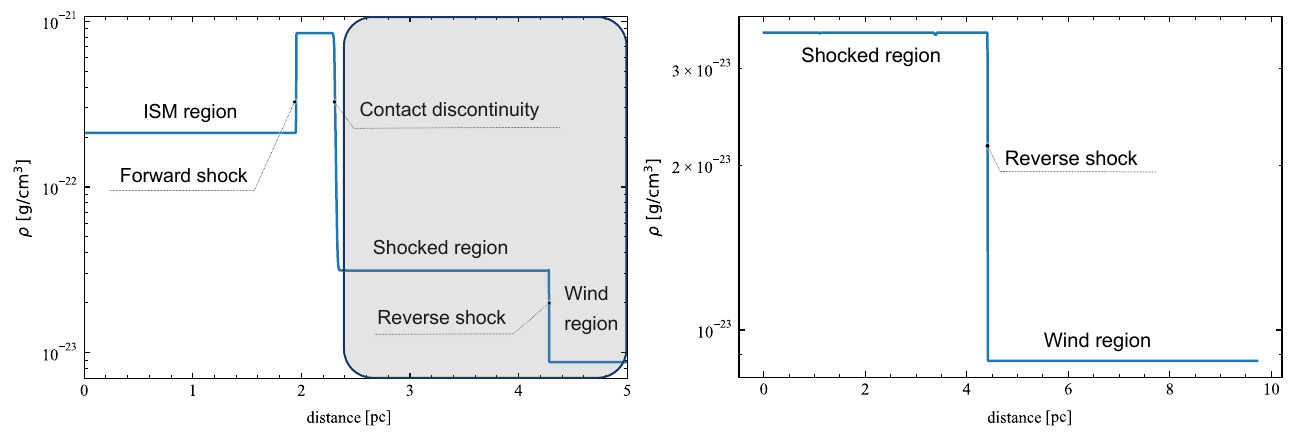}
\caption{\justifying 
Left: Density distribution of the global wind–ISM interaction reproduced by one-dimensional MHD simulations. 
Right: Extracted density profile around the reverse shock highlighted as grey region in Left panel, focusing on the local region used for particle acceleration analysis.}
\label{fig:shock_structure}
\end{figure*}

\subsubsection{Global Fluid Initial Conditions}

The wind launched from a supermassive black hole, including UFOs, is expected to form a bubble structure similar to stellar winds, as suggested by both theory and observations (see Fig.~\ref{fig:shock_structure} left) \citep{2012ASPC..460..235Z,2018MNRAS.474.3673R,2018ApJ...867...88R,2018ApJ...858....9L,2019A&A...628A.118B,2023ApJ...951..146M}. We performed one-dimensional MHD simulations of a Riemann problem, where a stationary homogeneous ISM is placed on the left side of $x=0$ and a leftward-propagating wind is injected on the right side. The time evolution of this system generates a shock structure consisting of a reverse shock, contact discontinuity, and forward shock. The resulting density distribution of the bubble is shown in Fig.~\ref{fig:shock_structure} (left). The detailed quantative conditions for the left and right states of the Riemann problem are given in Sec. \ref{subsec:wind_region} and \ref{subsec:ISM_region}. Solving Eqs. \eqref{eq:CR_evolution_equation1} and \eqref{eq:CR_evolution_equation2} for the whole system is computationally expensive. To reduce it, we extract only the local region around the reverse shock and solve Eqs. \eqref{eq:CR_evolution_equation1} and \eqref{eq:CR_evolution_equation2} in couple with the Bell-MHD Eqs. \eqref{eq:continuity_equation_Bell}-\eqref{eq:definition_energy_density} (see, right panel of Fig. \ref{fig:shock_structure}).

\subsubsection{Region Around the Reverse Shock}

The simulation domain is set to $L_{\text{box}} = 10~\text{pc}$ with a runtime of $t_{\text{final}} = 100~\text{yr}$.\footnote{For the fiducial model with $B_0=10^{-5}~\mathrm{G}$, the box size is increased to $L_{\text{box}} = 15~\text{pc}$ to prevent the anisotropic CR current (Eq.~\eqref{eq:return_current_Bell}) from reaching the boundary.} Typical wind velocities are $v_{\text{wind}} \sim 0.1c$–$0.4c$ \citep{2003MNRAS.345..705P,2005A&A...442..461D,2006ApJ...646..783M,2006AN....327.1012C,2007ApJ...670..978B,2009A&A...504..401C,2009ApJ...701..493R,2010A&A...521A..57T,2015MNRAS.451.4169G,2015Sci...347..860N,2019MNRAS.482.5316M,2021NatAs...5...13L,2025Natur.641.1132X}. This setup corresponds to the propagation timescale of a UFO traveling about 1 pc from the central black hole. Observationally, UFOs are typically detected at distances $<10^{18}~\text{cm}$ \citep{2012MNRAS.422L...1T,2015ARA&A..53..115K,2019GReGr..51...65Z}, placing the 1 pc scale within the observed distribution range.

The wind and ISM parameters are summarized in Tab.~\ref{tab:Model_Parameters_UFO}. The mass outflow rate is set to $\dot{M} \sim 0.01$–$1~M_{\odot}/\mathrm{yr}$ \citep{2012MNRAS.422L...1T,2015MNRAS.451.4169G,2019ApJ...871..156M,2021NatAs...5...13L}.

\subsubsection{Wind Region}
\label{subsec:wind_region}
The wind density is expressed as follows:
\begin{equation}
    \begin{aligned}
        \rho_{\text{wind}}(r) &= \frac{\dot{M}}{4\pi r^2 v_{\text{wind}}} \\
        &= 8.8\times10^{-24} ~\text{g}~\text{cm}^{-3}\\
        &\times\left(\frac{r}{1~\text{pc}}\right)^{-2} 
        \left(\frac{v_{\text{wind}}}{0.2c}\right)^{-1} 
        \left(\frac{\dot{M}}{0.1~M_{\odot}/\text{yr}}\right).
    \end{aligned}
    \label{eq:rho_wind_initial}
\end{equation}
The sound speed is defined as
\begin{equation}
    c_{\text{s}}\equiv\sqrt{\gamma\frac{P}{\rho}}.
    \label{eq:acoustic_wave_UFO}
\end{equation}
Using this, the Mach number is given by
\begin{equation}
    \mathcal{M}_{\text{wind}}\equiv\frac{v_{\text{wind}}}{c_{\text{s, wind}}}.
\end{equation}
The wind pressure can then be expressed as
\begin{equation}
    \begin{aligned}
        P_{\text{wind}} &= \rho_{\text{wind}}\frac{1}{\gamma} \left(\frac{v_{\text{wind}}}{\mathcal{M}_{\text{wind}}}\right)^2 \\
        &= 4.7\times10^{-7}~\text{dyn/cm}^{2} ~\left(\frac{r}{1~\text{pc}}\right)^{-2} 
        \left(\frac{\mathcal{M}_{\mathrm{wind}}}{20}\right)^{-2}\\
        &\quad \times 
        \left(\frac{\gamma}{5/3}\right)^{-1}
        \left(\frac{v_{\text{wind}}}{0.2c}\right) 
        \left(\frac{\dot{M}}{0.1~M_{\odot}/\text{yr}}\right).
    \end{aligned}
    \label{eq:P_wind_initial}
\end{equation}

\subsubsection{ISM Region}
\label{subsec:ISM_region}
For the ISM region, we assume a stationary homogeneous medium, representative of the narrow line region where relatively low-density gas is observed\footnote{AGN outflows including UFOs are suggested to form shocks in the narrow line region \citep{2021PASJ...73.1152J,2024ApJ...960...41M}.}. The density and temperature of narrow line region, estimated from emission-line diagnostics, are typically $n_{\text{ISM}} \sim 10^2$–$10^4/\text{cm}^3$ and $T_{\text{ISM}} \sim 1.0$–$2.5 \times 10^4~\text{K}$ \citep{1978ApJ...223...56K} (see Sec. 6 of \citet{1997iagn.book.....P}). The ISM density is given by
\begin{equation}
    \begin{aligned}
        &\rho_{\text{ISM}} = m_{\text{g}}n_{\text{ISM}} \\
        &= 2.1\times10^{-22}~\text{g/cm}^3  
        \left(\frac{m_{\text{g}}/m_{\text{p}}}{1.27}\right)
        \left(\frac{n_{\text{ISM}}}{10^2/\text{cm}^3}\right).
    \end{aligned}
    \label{eq:rho_ISM_initial}
\end{equation}
We consider three cases with $n_{\text{ISM}} = 1, 10^2, 10^4~\text{cm}^{-3}$ (Tab.~\ref{tab:Model_Parameters_UFO}). The ISM pressure is represented as
\begin{equation}
    \begin{aligned}
        &P_{\text{ISM}} = n_{\text{ISM}}k_{\text{B}}T_{\text{ISM}} \\
        &= 2.2\times10^{-10}~\text{dyn/cm}^{2}  
        \left(\frac{T_{\text{ISM}}}{1.6\times10^4~\text{K}}\right)
        \left(\frac{n_{\text{ISM}}}{10^2/\text{cm}^3}\right).
    \end{aligned}
    \label{eq:P_ISM_initial}
\end{equation}

\subsubsection{Initial Magnetic Fluctuations}
\label{subsec:initial_magnetic_field}
We superpose broadband circular-polarized Alfvén waves on to the background field in Eq. \eqref{eq:background_magnetic_field_UFO}. The fluctuation wavelengths are chosen such that the minimum wavelength is smaller than the maximum growth scale of the NRH instability (Eq.~\eqref{eq:maximum_growth_wavelength_Bell}), while the maximum wavelength is set to be equal to the box size:
\begin{equation}
    \lambda_{\text{min}}<\lambda_{\text{B}},\quad \lambda_{\text{max}}=L_{\text{box}}=3.0\times10^{19}~\text{cm}.
\end{equation}
The initial fluctuation spectrum follows a Kolmogorov form, $P_{k}(k)\propto k^{-\frac{5}{3}}$.

The fluctuation strength is defined as
\begin{equation}
    \xi_{B,\mathrm{ini}} \equiv \left\langle \frac{B_y^2 + B_z^2}{B_0^2} \right\rangle.
    \label{eq:definition_magnetic_fluctuation_strength}
\end{equation}
We adopt two cases: strong initial turbulence with $\xi_{B,\mathrm{ini}}=0.1$ and weak turbulence with $\xi_{B,\mathrm{ini}}=0.01$.

As shown in Tab.~\ref{tab:Model_Parameters_UFO}, we vary $B_0$, $\xi_{B,\mathrm{ini}}$, $n_{\text{ISM}}$, and the presence of $\mathrm{p}\gamma$ cooling to evaluate their impact on particle acceleration. Since observational constraints on $B_0$ in UFOs are limited, we explore a wide range of field strengths.

\subsubsection{Injection Rate}

The injection rate $\eta$ appearing in Eq.~\eqref{eq:injection_for_simulation} is set to $\eta=10^{-4}$ and $10^{-3}$, as summarized in Tab.~\ref{tab:Model_Parameters_UFO}. Varying $\eta$ changes the number density of CRs, thereby altering the growth efficiency of the NRH instability. Since UFO injection rates are poorly constrained observationally, we adopt $\eta=10^{-4}$–$10^{-3}$ as a representative range, guided by studies of DSA in SNRs \citep{2009A&A...499..191T,2012ApJ...759...12B}. This corresponds to cases where a few to tens of percent of the upstream kinetic energy is transferred into CR energy. 

\nishiura{Note that $\eta \sim 10^{-3}$ approaches the regime where nonlinear modification of the shock structure and the CR energy spectrum by CR pressure is expected \citep{1997ApJ...485..638M,1997ApJ...491..584M,2001RPPh...64..429M,2005MNRAS.361..907B,2008MNRAS.385.1946A,2009ApJ...694..951R,2013ApJ...775..130S}. However, in the present simulations, the MHD momentum and energy equations (Eqs.~\eqref{eq:EOM_Bell_x} and \eqref{eq:energy_equation_Bell}) do not yet include CR pressure terms, so these modifications cannot appear. Implementing this back-reaction will be an extension for future work.}

\subsection{Boundary Conditions}
For the MHD component, both the left and right boundaries are fixed. At the left boundary, we apply the wind parameters given by Eqs.~\eqref{eq:rho_wind_initial} and \eqref{eq:P_wind_initial}, with a fixed wind velocity of $0.2c$. At the right boundary, we impose the stationary ISM parameters described by Eqs.~\eqref{eq:rho_ISM_initial} and \eqref{eq:P_ISM_initial}.

For the CR component, in physical space, the isotropic part $F_0$ and anisotropic part $F_1$ of the CR distribution function are set to zero at both boundaries\footnote{The difference between imposing a zero boundary or a free boundary is negligible. Since the simulation domain is sufficiently large, the reverse shock remains far from the right boundary, and CRs never reach it within the simulation time.}. In momentum space, the computational domain is defined as
\begin{equation}
    p_{\text{min}}=10^{12}~\text{eV}~c^{-1},\quad p_{\text{max}}=10^{19}~\text{eV}~c^{-1},
\end{equation}
with the exception that for the $B_0=10^{-2}~\mathrm{G}$ model the minimum momentum is set to $p_{\text{min}}=10^{13}~\text{eV}~c^{-1}$.\footnote{At $p=10^{12}~\text{eV}~c^{-1}$, the diffusion length cannot be resolved with more than ten grid cells given the spatial resolution of Eq.~\eqref{eq:kuukan_bunkainou_UFO}. See Eq.~\eqref{eq:diffusion_length_UFO_bunkai} for details.} For $p_{\text{inj}}<p<p_{\text{min}}$, the isotropic component $F_0$ follows the DSA spectrum $f_0(x_{\text{sh}})\propto p^{-4}$ as expressed in Eq.~\eqref{eq:injection_for_simulation}. Outside the numerical domain of the momentum, the anisotropic component $F_1$ is set to zero.

\subsection{Numerical Resolution}

\subsubsection{Spatial Resolution}
The spatial resolution must satisfy three criteria simultaneously:
\begin{enumerate}[label=(\roman*)]
    \item The maximum growth wavelength $\lambda_{\text{NRH}}$ of the NRH instability must be resolved with at least 32 cells (see Fig.~21 of \citet{2021ApJ...922....7I}).
    \item The diffusion length $l_{\text{diff}}$ of the lowest-energy CRs ($p_{\text{min}}$) must be resolved with at least 10 cells (see Fig.~20 of \citet{2021ApJ...922....7I}).
    \item The shortest wavelength of the initial magnetic fluctuations, $\lambda_{\text{min}}$, must be sufficiently resolved to avoid numerical dissipation.
\end{enumerate}

The maximum growth wavelength $\lambda_{\text{NRH}}$ is given by \citet{2004MNRAS.353..550B} as
\begin{equation}
    \begin{aligned}
        \lambda_{\text{NRH}} &= \frac{B_0c}{j_{\text{CR}}} \\
        &= 4.9\times10^{14}~\text{cm}\\
        &\times\left(\frac{B_0}{10^{-5}~\text{G}}\right)
        \left(\frac{j_{\text{CR}}}{6.2\times10^{-10}~\text{esu s}^{-1}\text{cm}^{-2}}\right)^{-1},
    \end{aligned}
    \label{eq:maximum_growth_wavelength_Bell}
\end{equation}
where $j_{\text{CR}}$ is the CR current at the location where the NRH e-folding number $t_{\text{adv}}/t_{\text{NRH}}$ reaches its maximum. Test calculations estimate this value as $6.2\times10^{-10}~\text{esu s}^{-1}\text{cm}^{-2}$ for the fiducial model with $(B_0,~\eta,~\xi_{B,\mathrm{ini}})=(10^{-5}~\mathrm{G},~10^{-4},~0.1)$. The details of the NRH e-folding number are discussed in Sec.~\ref{subsubsec:small_magnetic_field_UFO}.

The diffusion length of the lowest-energy CRs is expressed as follows\footnote{The shock velocity in the UFO frame is estimated from the time evolution of the position where the fluid velocity discontinuously changes from $0~\mathrm{cm}/\mathrm{s}$ to beyond $2\times10^{9}~\mathrm{cm}/\mathrm{s}$.}:
\begin{equation}
    \begin{aligned}
        l_{\text{diff}} &\equiv \frac{D_\|}{v_{\text{sh}}}
        = \frac{4}{3\pi}\xi_{B}^{-1}\frac{cE_{\text{CR}}}{eB_0v_{\text{sh}}} \\
        &= 8.5\times10^{14}~\text{cm}
        \left(\frac{\xi_{B}}{0.1}\right)^{-1}
        \left(\frac{E_{\text{CR}}}{10^{12}~\text{eV}}\right) \\
        &\quad \times 
        \left(\frac{B_0}{10^{-4}~\text{G}}\right)^{-1}
        \left(\frac{v_{\text{sh}}}{5.0\times10^9~\text{cm}/\text{s}}\right)^{-1}.
    \end{aligned}
    \label{eq:diffusion_length_UFO_bunkai}
\end{equation}
Here, $v_{\text{sh}}$ is the shock velocity in the upstream rest frame.

To prevent numerical dissipation of the shortest Alfvén wavelengths, the spatial resolution must be chosen carefully. Previous studies analyzed the numerical dissipation of circularly polarized Alfvén waves using the second-order Roe flux method \citep{2008ApJS..178..137S}. Based on their results, the required number of numerical cells per wavelength $n_x$ for the shortest mode can be estimated as\footnote{
After \(v_{\text{A}}t_{\text{final}}/\lambda\) periods, the condition for the amplitude of an Alfvén wave with wavelength \(\lambda\) to remain above \(95\%\) can be expressed as
\begin{equation}
\begin{aligned}
\left\{1 - 0.20\left(\frac{8}{n_x}\right)^{2}\right\}^{\frac{v_{\text{A}}t_{\text{final}}}{5\lambda}} > 0.95,
\end{aligned}
\nonumber
\end{equation}
where the Alfvén speed \(v_{\text{A}}\propto B_0\) is defined by Eq.~\eqref{eq:definition_Alfven_speed_UFO}.
In this study, we choose $n_x$ according to the strength of the background magnetic field. Specifically, we set \(n_x = 16\) for \(B_{0} = 10^{-5}~\mathrm{G}\);
\(n_x = 64\) for \(B_{0} = 10^{-4}~\mathrm{G}\) and \(B_{0} = 10^{-3}~\mathrm{G}\);
and \(n_x = 256\) for \(B_{0} = 10^{-2}~\mathrm{G}\).
}
\begin{equation}
\begin{aligned}
n_x>224
&\left(\frac{\rho_{\text{wind}}}{8.78 \times 10^{-24}~\mathrm{g/cm}^3}\right)^{\frac{1}{4}}
\left(\frac{B_0}{10^{-2}~\mathrm{G}}\right)^{\frac{1}{2}} \\
&\times \left(\frac{t_{\text{final}}}{100~\mathrm{yr}}\right)^{\frac{1}{2}}
\left(\frac{\lambda}{10^{-3}~\mathrm{pc}}\right)^{-\frac{1}{2}}.
\end{aligned}
\label{eq:condition_magnetic_fluctuation_divide}
\end{equation}
Accordingly, the grid resolution is set to satisfy Eq.~\eqref{eq:condition_magnetic_fluctuation_divide} for the shortest initial wavelength $\lambda_{\text{min}}$. Combining all conditions, we adopt the following resolution:
\begin{equation}
    \Delta x=1.7\times10^{13}~\text{cm},\quad N_{\text{cell},x}=1,760,000.
    \label{eq:kuukan_bunkainou_UFO}
\end{equation}

\subsubsection{Momentum-Space Resolution}
The momentum space is divided uniformly in logarithmic intervals from $p_{\text{min}}$ to $p_{\text{max}}$. We use $N_{\text{cell},p} = 64$ cells, consistent with the convergence tests of \citet{2021ApJ...922....7I}.

\subsubsection{Time Step Requirement}
Since the evolution equations for the CR distribution functions, Eqs.~\eqref{eq:CR_evolution_equation1} and \eqref{eq:CR_evolution_equation2}, are hyperbolic, the time step must satisfy the Courant–Friedrichs–Lewy (CFL) condition, expressed as
\begin{equation}
    \Delta t\leq C_{\text{CFL}}\frac{\Delta x}{v_{\text{CR}}}\sim 3\times10^3~\text{s},
\end{equation}
where Eq.~\eqref{eq:kuukan_bunkainou_UFO} is used. The CR velocity is assumed to be $v_{\text{CR}} = c/\sqrt{3}$, corresponding to free streaming at relativistic speeds. We adopt $C_{\text{CFL}}=0.8$.

\section{Simulation Results}
\label{sec:result_UFO}
\subsection{Variation of Background Magnetic Field Strength}
\label{subsec:variation_of_magnetic_field_UFO}
\begin{figure*}[htbp]
  \centering
  \includegraphics[width=\textwidth]{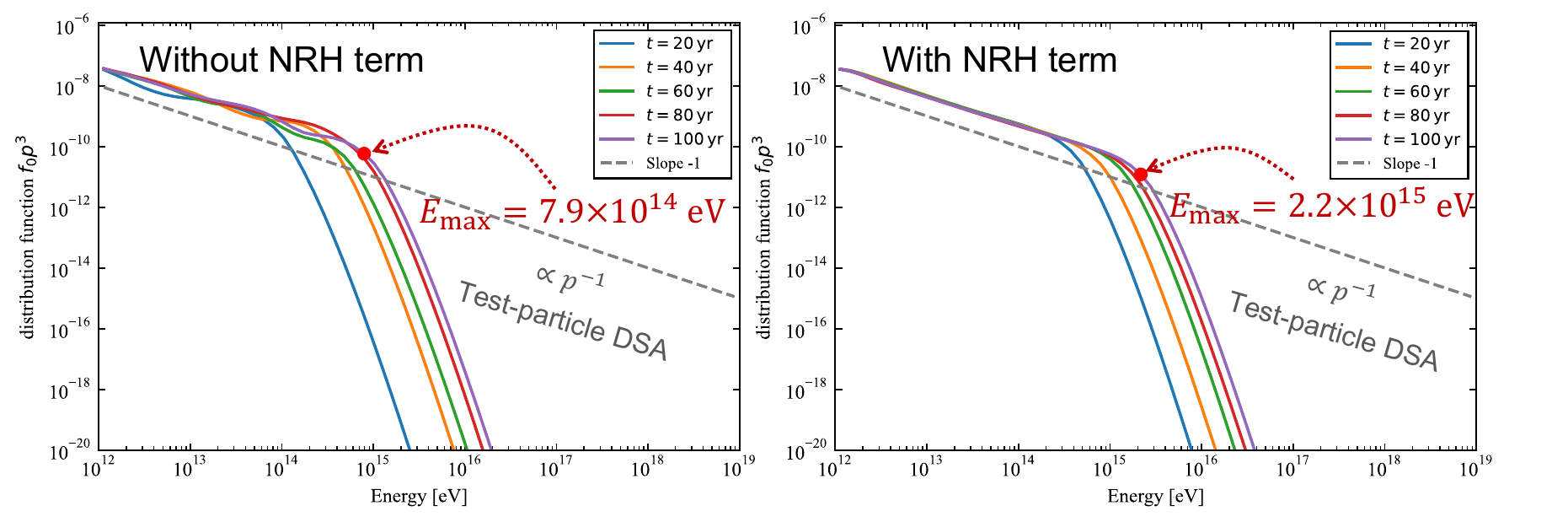}
  \caption{\justifying Energy spectra of the isotropic component of CRs in the Fiducial model \nishiura{with $(B_0,~\eta,~\xi_{B,\mathrm{ini}})=(10^{-5}~\mathrm{G},~10^{-4},~0.1)$} listed in Tab.~\ref{tab:Model_Parameters_UFO}. The left panel shows the case without NRH instability, while the right panel includes NRH instability. Blue, orange, green, red, and purple solid lines represent the spectra at 20, 40, 60, 80, and 100 yr after the start of the simulation, respectively. The gray dashed line denotes the power-law slope predicted by the \nishiura{test-particle DSA} solution. Red dots indicate the maximum acceleration energy, defined as the momentum at which the spectral index of $f_0p^3$ falls below $-2.1$.}
  \label{fig:CR_spectrum_fiducial}
\end{figure*}
\begin{table}[htbp]
  \centering
  \renewcommand{\arraystretch}{1.5}
  \caption{Maximum acceleration energy of CRs in the Fiducial model with $\xi_{B,\mathrm{ini}}=0.1$.}
  \label{tab:Maximum_Energy_Fiducial}
\begin{tabular}{lccc}
  \toprule
  $B_0$ [G] & No NRH & $\eta=10^{-4}$ & $\eta=10^{-3}$ \\
  \midrule
  $10^{-5}$ & $7.9\times10^{14}~\mathrm{eV}$ & $2.2\times10^{15}~\mathrm{eV}$ & $5.9\times10^{15}~\mathrm{eV}$ \\
  $10^{-4}$ & $5.9\times10^{15}~\mathrm{eV}$ & $9.8\times10^{15}~\mathrm{eV}$ & $4.5\times10^{16}~\mathrm{eV}$ \\
  $10^{-3}$ & $9.5\times10^{16}~\mathrm{eV}$ & $9.5\times10^{16}~\mathrm{eV}$ & $1.2\times10^{17}~\mathrm{eV}$ \\
  $10^{-2}$ & $1.0\times10^{18}~\mathrm{eV}$ & $8.4\times10^{17}~\mathrm{eV}$ & $9.1\times10^{17}~\mathrm{eV}$ \\
  \bottomrule
\end{tabular}
\end{table}
\begin{table*}[htbp]
  \centering
  \renewcommand{\arraystretch}{1.25}
  \caption{Comparison of the maximum acceleration energy $E_{\max}$ [$\mathrm{eV}$] in the Fiducial model with $\eta=10^{-4}$. Both cases with initial magnetic fluctuations of $\xi_{B,\mathrm{ini}}=0.1$ and $0.01$ are shown. The ratio in the last column shows the dependence on the initial fluctuation amplitude when NRH instability is included. \nishiura{Here $\mathcal{M}_{\mathrm{A}}$ denotes the Alfv\'en Mach number.}}
  \label{tab:Emax_compare_eta1e-4}
  \begin{tabular}{ccccccc}
    \toprule
    \multicolumn{1}{c}{$B_0$ [G]} &
    \multicolumn{1}{c}{\nishiura{$\mathcal{M}_{\mathrm{A}}$}} &
    \multicolumn{2}{c}{$\xi_{B,\mathrm{ini}}=0.1$} &
    \multicolumn{2}{c}{$\xi_{B,\mathrm{ini}}=0.01$} &
    \multicolumn{1}{c}{\parbox[c][2.8ex][c]{10em}{\centering Ratio with NRH\\ $(\xi_{B,\mathrm{ini}}{=}0.1)/(\xi_{B,\mathrm{ini}}{=}0.01)$}} \\
    \cmidrule(lr){3-4}\cmidrule(lr){5-6}
    & & No NRH & with NRH & No NRH & with NRH & \\
    \midrule
    $10^{-5}$ & $6.3\times10^{3}$ & $7.9\times10^{14}~\mathrm{eV}$ & $2.2\times10^{15}~\mathrm{eV}$ & $8.2\times10^{13}~\mathrm{eV}$ & $1.3\times10^{15}~\mathrm{eV}$ & $1.7$ \\
    $10^{-4}$ & $6.3\times10^{2}$ & $5.9\times10^{15}~\mathrm{eV}$ & $9.8\times10^{15}~\mathrm{eV}$ & $6.2\times10^{14}~\mathrm{eV}$ & $7.6\times10^{15}~\mathrm{eV}$ & $1.3$ \\
    $10^{-3}$ & $6.3\times10^{1}$ & $9.5\times10^{16}~\mathrm{eV}$ & $9.5\times10^{16}~\mathrm{eV}$ & $7.6\times10^{15}~\mathrm{eV}$ & $1.3\times10^{16}~\mathrm{eV}$ & $9.2$ \\
    $10^{-2}$ & $6.3$             & $1.0\times10^{18}~\mathrm{eV}$ & $8.4\times10^{17}~\mathrm{eV}$ & $7.4\times10^{16}~\mathrm{eV}$ & $9.5\times10^{16}~\mathrm{eV}$ & $8.8$ \\
    \bottomrule
  \end{tabular}
\end{table*}

\begin{table}[htbp]
  \centering
  \renewcommand{\arraystretch}{1.5}
  \caption{Magnetic energy fraction $\varepsilon_B$ upstream (up) and downstream (down) of the shock for $\eta=10^{-4}$. Results are shown for each $B_0$ and initial fluctuation amplitude $\xi_{B,\mathrm{ini}}$, with and without NRH instability.}
  \label{tab:epsB_up_down_summary}
  \begin{tabular}{lcccccc}
    \toprule
    $B_0$ [G] & $\xi_{B,\mathrm{ini}}$ & NRH & $E_{\max}$ [eV] & $\varepsilon_B^{\mathrm{up}}$ & $\varepsilon_B^{\mathrm{down}}$ \\
    \midrule
    $10^{-5}$ & $0.1$  & No  & $7.9\times10^{14}$ & $4.3\times10^{-9}$  & $3.3\times10^{-8}$ \\
    $10^{-5}$ & $0.1$  & Yes & $2.2\times10^{15}$ & $3.5\times10^{-8}$  & $6.3\times10^{-8}$ \\
    $10^{-5}$ & $0.01$ & No  & $8.2\times10^{13}$ & $4.3\times10^{-10}$ & $3.3\times10^{-9}$ \\
    $10^{-5}$ & $0.01$ & Yes & $1.3\times10^{15}$ & $1.9\times10^{-8}$  & $4.9\times10^{-8}$ \\
    \midrule
    $10^{-4}$ & $0.1$  & No  & $5.9\times10^{15}$ & $8.0\times10^{-7}$  & $6.0\times10^{-6}$ \\
    $10^{-4}$ & $0.1$  & Yes & $9.8\times10^{15}$ & $5.0\times10^{-7}$  & $5.2\times10^{-5}$ \\
    $10^{-4}$ & $0.01$ & No  & $6.2\times10^{14}$ & $8.0\times10^{-8}$  & $6.0\times10^{-7}$ \\
    $10^{-4}$ & $0.01$ & Yes & $7.6\times10^{15}$ & $1.1\times10^{-6}$  & $1.5\times10^{-4}$ \\
    \midrule
    $10^{-3}$ & $0.1$  & No  & $9.5\times10^{16}$ & $1.2\times10^{-5}$  & $1.6\times10^{-4}$ \\
    $10^{-3}$ & $0.1$  & Yes & $9.5\times10^{16}$ & $7.8\times10^{-6}$  & $4.6\times10^{-4}$ \\
    $10^{-3}$ & $0.01$ & No  & $7.6\times10^{15}$ & $1.2\times10^{-6}$  & $1.6\times10^{-5}$ \\
    $10^{-3}$ & $0.01$ & Yes & $1.3\times10^{16}$ & $2.3\times10^{-6}$  & $1.3\times10^{-3}$ \\
    \midrule
    $10^{-2}$ & $0.1$  & No  & $1.0\times10^{18}$ & $3.2\times10^{-3}$  & $4.0\times10^{-2}$ \\
    $10^{-2}$ & $0.1$  & Yes & $8.4\times10^{17}$ & $3.2\times10^{-3}$  & $4.0\times10^{-2}$ \\
    $10^{-2}$ & $0.01$ & No  & $7.4\times10^{16}$ & $8.8\times10^{-5}$  & $2.4\times10^{-3}$ \\
    $10^{-2}$ & $0.01$ & Yes & $9.5\times10^{16}$ & $3.9\times10^{-4}$  & $3.6\times10^{-3}$ \\    
    \bottomrule
  \end{tabular}
\end{table}
In this section, we discuss the results for the Fiducial models in Tab.~\ref{tab:Model_Parameters_UFO}, which assumes a mass outflow rate of $0.1~M_{\odot}/\mathrm{yr}$ and an ISM density of $10^2~\mathrm{cm}^{-3}$, with $B_0$ varied from $10^{-5}$ to $10^{-2}~\mathrm{G}$.  
Tab.~\ref{tab:Maximum_Energy_Fiducial} summarizes the maximum CR energy for each $B_0$ with $\xi_{B,\mathrm{ini}}=0.1$, comparing cases with and without NRH instability and for different injection rates.  
Tab.~\ref{tab:Emax_compare_eta1e-4} focuses on $\eta=10^{-4}$, contrasting cases with initial magnetic fluctuations $\xi_{B,\mathrm{ini}}=0.1$ and $\xi_{B,\mathrm{ini}}=0.01$. \nishiura{
Here, we define the Alfvén Mach number as
\begin{equation}
\begin{aligned}
&\mathcal{M}_{\mathrm{A}}
\equiv \frac{v_{\mathrm{wind}}}{v_{\mathrm{A}}} \\
&= 6.3\times10^{3}
\left(\frac{\rho_{\mathrm{wind}}}{8.78\times10^{-24}~\mathrm{g}\,\mathrm{cm}^{-3}}\right)^{1/2}
\left(\frac{B_{0}}{10^{-5}~\mathrm{G}}\right)^{-1}.
\end{aligned}
\label{eq:alfven_mach_number}
\end{equation}
}These results show that for $B_0=10^{-5}$ and $10^{-4}~\mathrm{G}$, the maximum energy converges to nearly the same value regardless of $\xi_{B,\mathrm{ini}}$ owing to NRH instability (see Sec. \ref{subsubsec:small_magnetic_field_UFO}).  
In contrast, for $B_0>10^{-3}~\mathrm{G}$ the NRH instability becomes inefficient, and the maximum energy is determined by $\xi_{B,\mathrm{ini}}$ (see Sec.~\ref{subsubsec:large_magnetic_field_UFO}). \nishiura{This transition corresponds to  $\mathcal{M}_{\mathrm{A}}\sim10^{2}$.}

Tab.~\ref{tab:epsB_up_down_summary} further presents the magnetic energy fraction $\varepsilon_B$, defined as the ratio of magnetic energy to upstream kinetic energy, with $\varepsilon_B^{\mathrm{up}}$ and $\varepsilon_B^{\mathrm{down}}$ referring to the upstream and downstream regions, respectively:
\begin{equation}
\begin{aligned}
\varepsilon_B^{\mathrm{up}} &\equiv 
 \frac{\left\langle \dfrac{\delta B^2}{8\pi} \right\rangle_{[0,\,l_d]}}
 {\tfrac{1}{2}\rho_{\mathrm{wind}}\,v_{\mathrm{wind}}^2}, \\[1ex]
\varepsilon_B^{\mathrm{down}} &\equiv 
 \frac{\left\langle \dfrac{\delta B^2}{8\pi} \right\rangle_{[-l_d,\,0]}}
 {\tfrac{1}{2}\rho_{\mathrm{wind}}\,v_{\mathrm{wind}}^2}.
\end{aligned}
\end{equation}
Here, $l_{\mathrm{d}}\equiv\frac{4}{3\pi}\frac{cE_\mathrm{max}(\mathrm{No~NRH})}{eB_0v_{\mathrm{sh}}}$ represents the diffusion length estimated from the maximum energy in the absence of NRH instability (Tab.~\ref{tab:Maximum_Energy_Fiducial}), assuming $\xi_{B,\mathrm{ini}}=1$.\footnote{This choice corresponds to the characteristic scale where downstream magnetic fluctuations remain nearly constant due to shock compression, and provides a consistent averaging length across different $B_0$.}  
In all cases, $\varepsilon_B$ is smaller upstream than downstream, indicating that magnetic amplification in the upstream region controls the overall efficiency of particle acceleration as expected.

\subsubsection{\nishiura{Test-particle }prediction from diffusive shock acceleration}

In the framework of the DSA model, the energy spectrum of CRs can be analytically derived. In the limit of infinite Mach number, the isotropic distribution function is expressed as \citep{1978MNRAS.182..147B,1978ApJ...221L..29B,1983RPPh...46..973D,1987PhR...154....1B}  
\begin{equation}
f_0(p) \propto p^{-4}.
\end{equation} 

The maximum acceleration energy can also be estimated within the DSA framework. The acceleration timescale can be estimated by using Eq. \eqref{eq:definition_magnetic_fluctuation_strength}, 
\begin{equation}
t_{\mathrm{acc}} \equiv \frac{D_{\|}}{v_{\mathrm{sh}}^2}
= \frac{4}{3 \pi} \frac{c E_{\mathrm{CR}}}{e B_0 v_{\mathrm{sh}}^2}\xi_{B}^{-1}.
\label{eq:acceleration_time_UFO}
\end{equation}  
By equating $t_{\mathrm{acc}}$ with the time $t$, the maximum energy is obtained as  
\begin{equation}
\begin{aligned}
E_{\text{max}} &= \frac{3\pi}{4}\frac{e B_0 v_{\mathrm{sh}}^2}{c}t\xi_{B} \\
&\sim 1.8 \times 10^{15}~\mathrm{eV}
\left(\frac{B_0}{10^{-5}~ \mathrm{G}}\right)\\
&\times\left(\frac{v_{\mathrm{sh}}}{5.0\times10^9~\text{cm/s}}\right)^2
\left(\frac{\xi_{B}}{0.1}\right)
\left(\frac{t}{100~ \mathrm{yr}}\right),
\end{aligned}
\label{eq:order_estimation_maximum_energy_CR_Bell}
\end{equation}
which will be used in the following discussions.
\subsubsection{Case without NRH instability}

We first examine the simulation results without the NRH term and compare them with the \nishiura{test-particle} predictions of DSA. Fig.~\ref{fig:CR_spectrum_fiducial} shows the CR energy spectra for $B_0=10^{-5}~\mathrm{G}$, with and without NRH instability. In both cases, the isotropic distribution function follows the \nishiura{test-particle DSA} prediction $p^{-4}$ with good agreement.  

The maximum CR energies without the NRH term are consistent with Eq. \eqref{eq:order_estimation_maximum_energy_CR_Bell}. As shown in Tab.~\ref{tab:Maximum_Energy_Fiducial}, the simulation results match the order-of-magnitude estimate of Eq.~\eqref{eq:order_estimation_maximum_energy_CR_Bell} within a factor of two. In addition, Tab.~\ref{tab:Emax_compare_eta1e-4} shows that reducing the initial fluctuation amplitude $\xi_{B,\mathrm{ini}}$ by an order of magnitude lowers the maximum energy by a similar factor, consistent with Eq.~\eqref{eq:order_estimation_maximum_energy_CR_Bell}.

\subsubsection{Case of weak background magnetic field with NRH instability ($B_0=10^{-5},\,10^{-4}~\mathrm{G}$)}
\label{subsubsec:small_magnetic_field_UFO}

The key result for $B_0=10^{-5},\,10^{-4}~\mathrm{G}$ is that, once NRH instability is included, $E_{\max}$ converges to nearly a same value for a given $B_0$, even when the initial fluctuation amplitude $\xi_{B,\mathrm{ini}}$ is varied (Tab.~\ref{tab:Emax_compare_eta1e-4}). In particular, when $\eta=10^{-4}$, reducing $\xi_{B,\mathrm{ini}}$ from $0.1$ to $0.01$ changes $E_{\max}$ by only factor 2 or less. This behavior indicates that even when the initial field amplitude is small, the growth and saturation of NRH instability render the final value of $E_{\max}$ insensitive to $\xi_{B,\mathrm{ini}}$.

The same convergence trend is evident in the magnetic energy fraction $\varepsilon_B$, as summarized in Tab.~\ref{tab:epsB_up_down_summary}. When NRH instability is included, $\varepsilon_B$ converges to similar values for $\xi_{B,\mathrm{ini}}=0.1$ and $\xi_{B,\mathrm{ini}}=0.01$. Moreover, in almost all cases, $\varepsilon_B^{\mathrm{up}}$ is smaller than $\varepsilon_B^{\mathrm{down}}$, implying that magnetic-field amplification upstream regulates the efficiency of particle acceleration.

\begin{figure*}
\centering
\includegraphics[width=\textwidth]{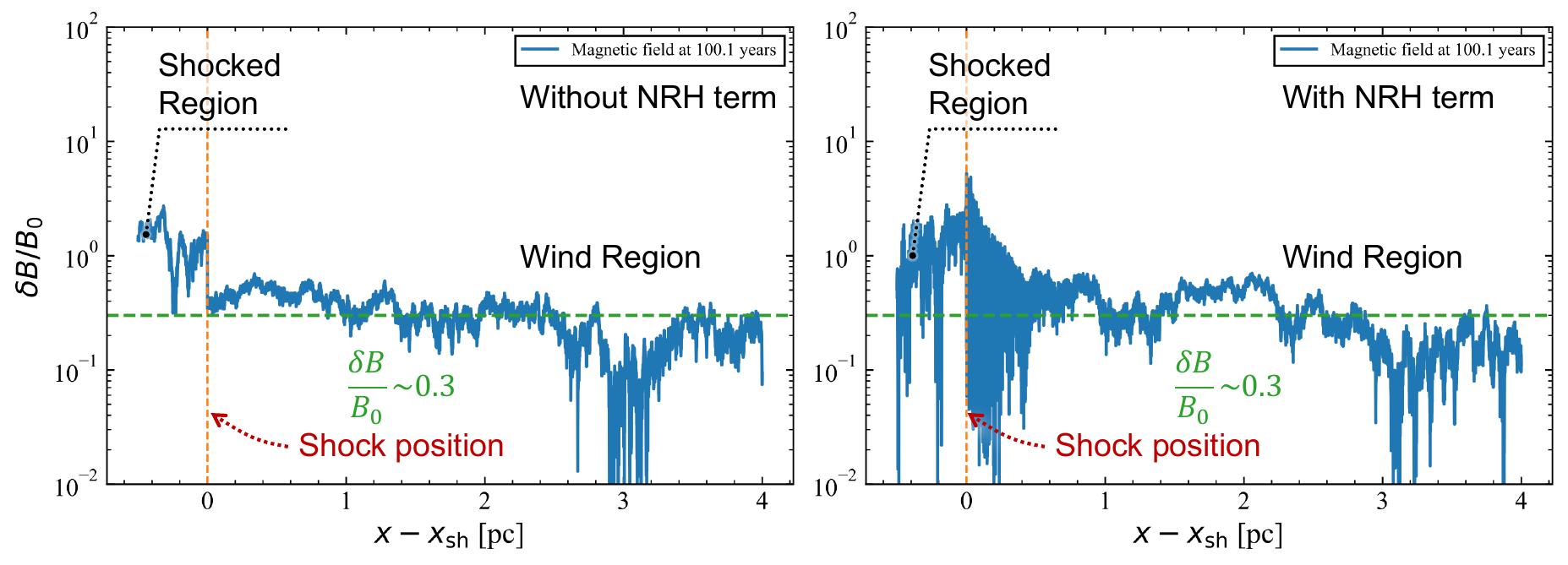}
\caption{\justifying 
Spatial profile of magnetic fluctuations $\delta B/B_0$ in the fiducial model with $(B_0,~\eta,~\xi_{B,\mathrm{ini}})=(10^{-5}~\mathrm{G},~10^{-4},~0.1)$ at $t=100~\mathrm{yr}$. Left: without NRH instability. Right: with NRH instability. The horizontal axis denotes the distance from the shock front, normalized so that the shock position is $x=0$ (in pc). The vertical axis shows $\delta B/B_0$. The orange dashed line marks the shock position, while the green dot-dashed line represents the mean amplitude of the initial fluctuations ($\delta B/B_0 \sim 0.3$).}
\label{fig:magnetic_fluctuation_from_shock_10-5}
\end{figure*}

\begin{figure*}
\centering
\includegraphics[width=\textwidth]{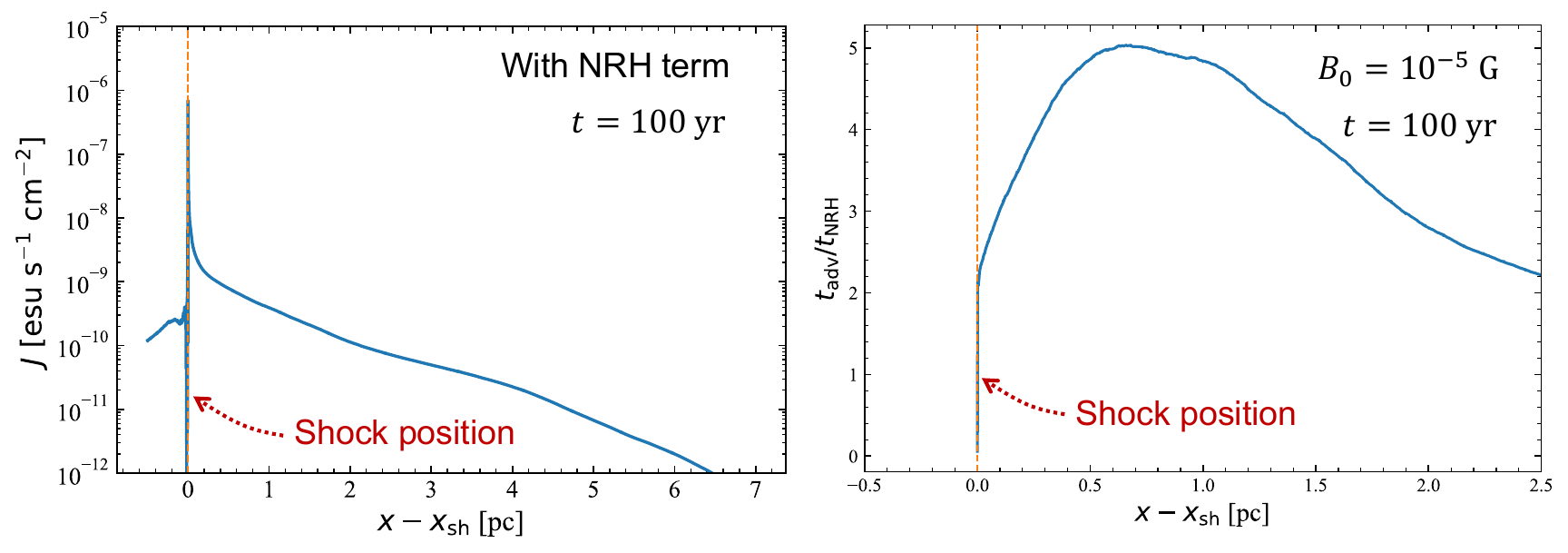}
\caption{\justifying 
Upstream profiles for the fiducial model with $(B_0,~\eta,~\xi_{B,\mathrm{ini}})=(10^{-5}~\mathrm{G},~10^{-4},~0.1)$ at $t=100~\mathrm{yr}$. The horizontal axis indicates the distance from the shock position (in pc), with the orange dashed line marking the shock. Left: current density carried by the CR anisotropic component $j_{\text{CR}}$ (esu s$^{-1}$ cm$^{-2}$). Right: e-folding number of NRH instability $t_{\mathrm{adv}}/t_{\mathrm{NRH}}$ (dimensionless), evaluated using Eq.~\eqref{eq:NRH_linear_timescale} and Eq.~\eqref{eq:definition_advection_time_Bell}. See also Tab.~\ref{tab:NRH_timescale_compare}.}
\label{fig:CR_density_from_shock_10-5}
\end{figure*}

Fig.~\ref{fig:magnetic_fluctuation_from_shock_10-5} shows the spatial distribution of $\delta B/B_0$ upstream of the shock at $t_{\text{final}}=100~\mathrm{yr}$ for $(B_0,~\eta,~\xi_{B,\mathrm{ini}})=(10^{-5}~\mathrm{G},~10^{-4},~0.1)$. Without NRH instability (left), the fluctuations remain close to the initial value, while with NRH instability (right), the magnetic field is amplified by nearly an order of magnitude near the shock front. The amplification decreases with distance from the shock, as clarified later in the discussion of Fig.~\ref{fig:CR_density_from_shock_10-5}. A similar trend is obtained for $B_0=10^{-4}~\mathrm{G}$. Meanwhile, in the downstream region, magnetic turbulence is enhanced by shock compression in both cases, with and without the NRH term.

Fig.~\ref{fig:CR_density_from_shock_10-5} presents the upstream profiles of the CR current density $j_{\text{CR}}$ (left), defined by Eq.~\eqref{eq:return_current_Bell}, and the e-folding number of NRH instability $t_{\text{adv}}/t_{\text{NRH}}$ (right). The e-folding number peaks in a finite region ahead of the shock, and Fig.~\ref{fig:magnetic_fluctuation_from_shock_10-5} confirms that magnetic-field amplification occurs predominantly inside this region. The inverse of the linear growth rate of the NRH instability (growth timescale) is represented by \citep{2004MNRAS.353..550B}
\begin{equation}
\begin{aligned}
   t_{\text {NRH }}&=\sqrt{\frac{\rho}{\pi}} \frac{c}{j_{\text{CR}}}\\
   &\sim 2.6 ~\mathrm{yr}~\left(\frac{\rho_{\text {wind }}}{8.78 \times 10^{-24} \mathrm{~g} / \mathrm{cm}^3}\right)^{\frac{1}{2}}\\
   &\times\left(\frac{j_{\text{CR}}}{6.2\times10^{-10}~ \mathrm{esu~s}^{-1} \mathrm{~cm}^{-2}}\right)^{-1},
\end{aligned}
\label{eq:NRH_linear_timescale}
\end{equation}
while the advection time before the shock overtakes that point is expressed as
\begin{equation}
\begin{aligned}
t_{\text{adv}}&\equiv \frac{x-x_{\text{sh}}}{v_{\text{sh}}}\\
&\sim 13~{\text{yr}}~\left(\frac{v_{\mathrm{sh}}}{5.0\times10^9~\text{cm/s}}\right)^{-1}
\left(\frac{x-x_{\text{sh}}}{6.6\times10^{-1}~\text{pc}}\right).
\end{aligned}
\label{eq:definition_advection_time_Bell}
\end{equation}
Magnetic-field amplification proceeds efficiently in regions where $t_{\text{NRH}} \ll t_{\text{adv}}$. In the case of $B_0=10^{-5}~\mathrm{G}$, the e-folding number reaches a maximum of $\sim5.0$ at $x-x_{\text{sh}}\sim6.6\times10^{-1}~\text{pc}$, and significant amplification occurs inside this location\footnote{The e-folding number in Fig.~\ref{fig:CR_density_from_shock_10-5} (right) represents a local measure of how many times the instability can grow before the shock arrival at each position. The actual number is larger inside the peak because growth initiated farther upstream continues to accumulate over time.}. A comparison of the maximum e-folding numbers for weak and strong background magnetic field cases is provided in Tab.~\ref{tab:NRH_timescale_compare}. The case of strong background fields ($B_0=10^{-2}~\mathrm{G}$) is discussed in more depth in Sec.~\ref{subsubsec:large_magnetic_field_UFO}.

\begin{table}[htbp]
\centering
\renewcommand{\arraystretch}{1.3}
\caption{Comparison of the maximum e-folding number of NRH instability $(t_{\text{adv}}/t_{\text{NRH}})_{\text{max}}$, its location $x-x_{\text{sh}}$, and the CR current density $j_{\text{CR}}$ at the maximum for different values of $B_0$.}
\label{tab:NRH_timescale_compare}
\begin{tabular}{cccc}
\hline
$B_0$ [G] & $(t_{\text{adv}}/t_{\text{NRH}})_{\text{max}}$ & $x-x_{\text{sh}}$ [pc] & $j_{\text{CR}}$ [esu s$^{-1}$ cm$^{-2}$] \\
\hline
$10^{-5}$ & $5.0$ & $6.6\times10^{-1}$ & $6.2\times10^{-10}$ \\
$10^{-2}$ & $1.4\times10^{-2}$ & $1.1\times10^{-1}$ & $1.1\times10^{-11}$ \\
\hline
\end{tabular}
\end{table}

\begin{figure*}
\centering
\includegraphics[width=\textwidth]{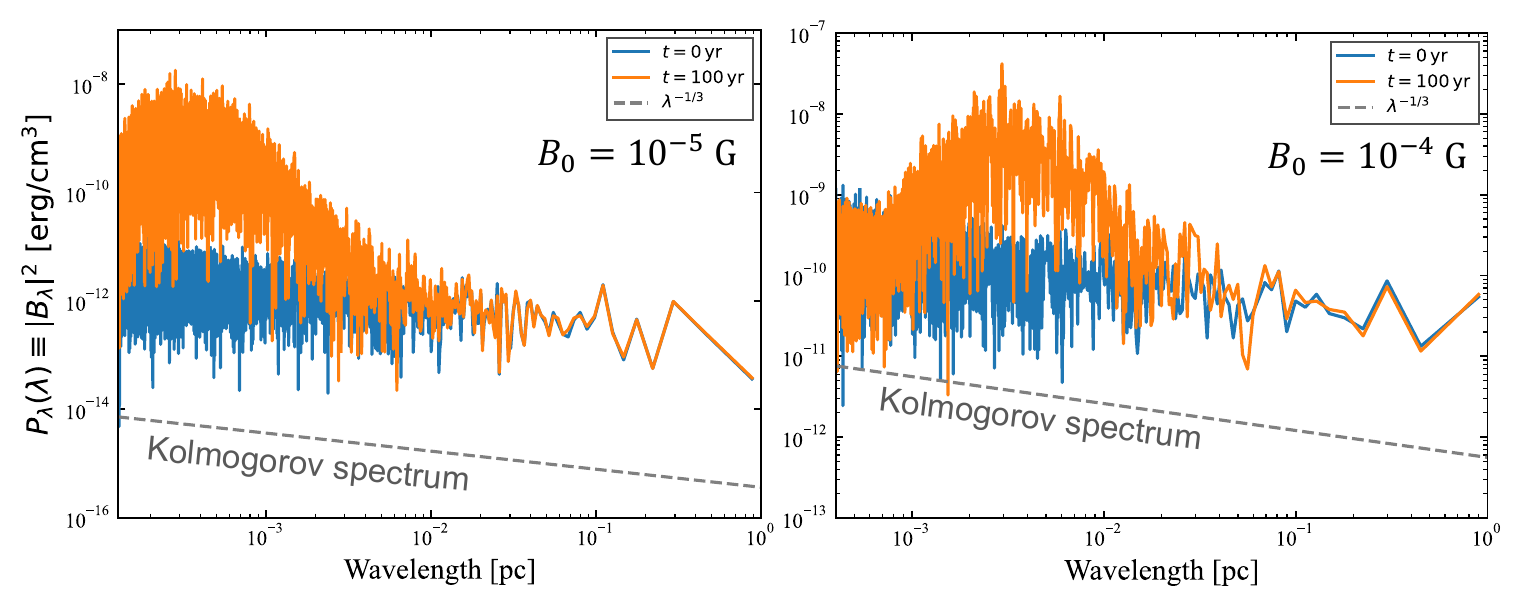}
\caption{\justifying 
Magnetic fluctuation wavelength spectra in the fiducial model (Tab.~\ref{tab:Model_Parameters_UFO}) for $(\eta,~\xi_{B,\mathrm{ini}})=(10^{-4},~0.1)$ including NRH instability. 
The left panel shows the case with $B_0=10^{-5}~\mathrm{G}$ and the right panel with $B_0=10^{-4}~\mathrm{G}$. 
In both cases, spectra are extracted at $t=100~\mathrm{yr}$ from the upstream region spanning $0.1$--$1~\mathrm{pc}$ ahead of the shock. 
The blue line indicates the initial Kolmogorov spectrum, the orange line shows the spectrum at $t=100~\mathrm{yr}$, and the gray dashed line represents the slope of a Kolmogorov spectrum $P_\lambda(\lambda)\propto \lambda^{-1/3}$, as defined in Eq.~\eqref{eq:magnetic_power_spectrum_UFO}.}
\label{fig:magnetic_spectrum_10-5}
\end{figure*}
Fig.~\ref{fig:magnetic_spectrum_10-5} presents the wavelength spectra of magnetic fluctuations extracted from the upstream region ($0.1$--$1~\mathrm{pc}$ ahead of the shock) in the fiducial model with $(\eta,~\xi_{B,\mathrm{ini}})=(10^{-4},~0.1)$, including the NRH instability. The left panel corresponds to $B_0=10^{-5}~\mathrm{G}$, and the right panel to $B_0=10^{-4}~\mathrm{G}$. In this analysis, the Fourier components $\tilde{B}_i(k)$, the power spectrum $P_k(k)$, and the wavelength spectrum $P_\lambda(\lambda)$ are defined as follows,
\begin{equation}
\begin{aligned}
\tilde{B}_i(k) & = \int B_i~\mathrm{e}^{-2 \pi \mathrm{i} k x} \dd x \quad (i=y, z), \\
P_k(k) & \equiv \left|\tilde{B}_y\right|^2 + \left|\tilde{B}_z\right|^2, \\
P_\lambda(\lambda) & \equiv P_k(k)\left|\frac{\dd k}{\dd \lambda}\right|.
\end{aligned}
\label{eq:magnetic_power_spectrum_UFO}
\end{equation}
For a Kolmogorov spectrum ($\propto k^{-3/5}$), the wavelength spectrum scales as $P_\lambda(\lambda)\propto \lambda^{-1/3}$.  

In the $B_0=10^{-5}~\mathrm{G}$ case, the spectrum at the final time peaks at $\lambda \sim 2.8\times10^{-4}~\mathrm{pc}$, consistent within a factor of two with the analytically expected NRH maximum growth scale $\lambda_{\text{NRH}}\sim1.6\times10^{-4}~\mathrm{pc}$ (Eq.~\eqref{eq:maximum_growth_wavelength_Bell}). The final spectrum shows rapid amplification from the shortest scales up to $\lambda_{\text{NRH}}$, followed by a gradual decline toward longer wavelengths. This behavior is explained by the NRH linear growth rate, expressed as \citep{2004MNRAS.353..550B}
\begin{equation}
t_{\text{NRH}}^{-1}(k) = \sqrt{k\left(\frac{B_0 j_{\text{CR}}}{c \rho} - k v_{\text{A}}^2\right)},
\end{equation}
where the Alfv\'en speed is defined by
\begin{equation}
v_{\text{A}} \equiv \frac{B_0}{\sqrt{4\pi \rho}}.
\label{eq:definition_Alfven_speed_UFO}
\end{equation}
Accordingly, the minimum unstable wavelength is represented by
\begin{equation}
\lambda_{\text{NRH}}^{\min} = \frac{B_0 c}{2 j_{\text{CR}}} = \frac{1}{2}\lambda_{\text{NRH}}.
\end{equation}
Thus, instability grows rapidly from $\lambda_{\text{NRH}}^{\min}$, reaches maximum growth at $\lambda_{\text{NRH}}$, and decreases toward longer wavelengths gradually.  

For $B_0=10^{-4}~\mathrm{G}$ (right panel), the NRH instability remains effective, producing amplification near the maximum growth scale. However, the spectral peak shifts to longer wavelengths, reaching $\sim2.9\times10^{-3}~\mathrm{pc}$. This shift is consistent with the scaling $\lambda_{\text{NRH}}\propto B_0/j_{\text{CR}}$, derived from Eq.~\eqref{eq:maximum_growth_wavelength_Bell} and the growth rate above. Since the spatial structure of $j_{\text{CR}}$ does not change significantly with $B_0$, increasing the background field by a factor of ten results in a nearly proportional increase of the peak wavelength.

\subsubsection{Cases with Strong Background Magnetic Fields ($B_0=10^{-3},~10^{-2}~\mathrm{G}$)}
\label{subsubsec:large_magnetic_field_UFO}
\begin{figure*}
\centering
\includegraphics[width=\textwidth]{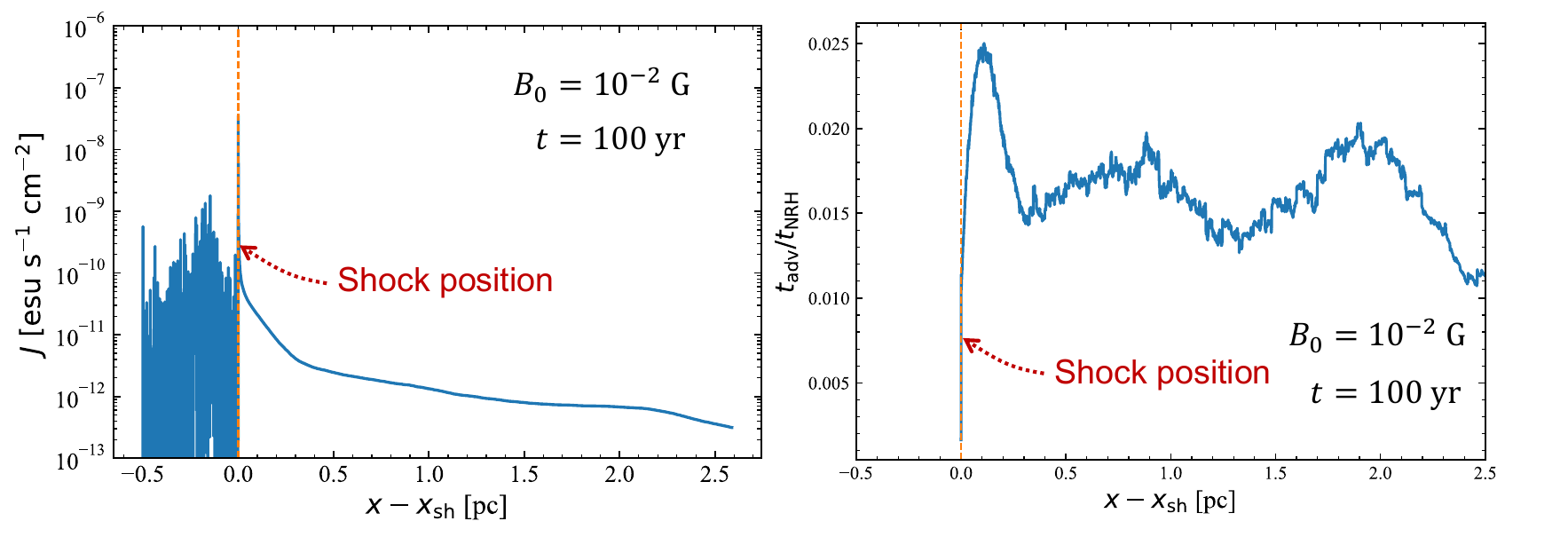}
\caption{\justifying Upstream profiles for the fiducial model with $(B_0,~\eta,~\xi_{B,\mathrm{ini}})=(10^{-2}~\mathrm{G},~10^{-4},~0.1)$ at $t=100~\mathrm{yr}$. The horizontal axis denotes the distance from the shock position (in pc), with the orange dashed line marking the shock. Left: spatial profile of the CR current density $j_{\text{CR}}$ generated by the anisotropic CR component (in esu s$^{-1}$ cm$^{-2}$). Right: spatial distribution of the NRH instability e-folding number $t_{\mathrm{adv}}/t_{\mathrm{NRH}}$ (dimensionless), evaluated using Eq.~\eqref{eq:NRH_linear_timescale} and Eq.~\eqref{eq:definition_advection_time_Bell}. See also Tab.~\ref{tab:NRH_timescale_compare}.}
\label{fig:CR_current_density_magnetic_field_difference}
\end{figure*}

\begin{figure*}
\centering
\includegraphics[width=\textwidth]{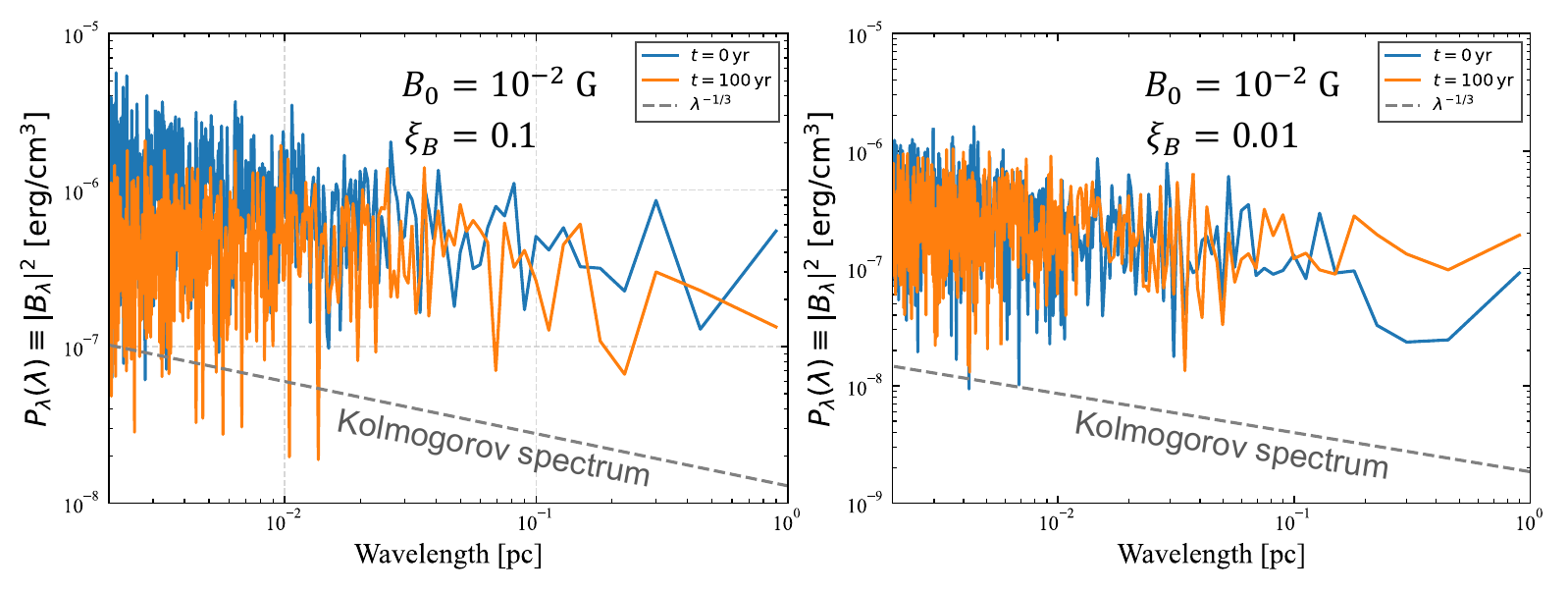}
\caption{\justifying Wavelength energy spectra of magnetic fluctuations in the fiducial model with $(B_0,~\eta)=(10^{-2}~\mathrm{G},~10^{-4})$ including NRH instability. The left panel shows the case with initial fluctuation amplitude $\xi_{B,\mathrm{ini}}=0.1$, and the right panel shows $\xi_{B,\mathrm{ini}}=0.01$. Spectra are calculated at the final simulation time $t=100~\mathrm{yr}$, extracted from the upstream region $0.1$--$1~\mathrm{pc}$ ahead of the shock. Blue curves indicate the initial Kolmogorov spectrum, while orange curves show the spectra at $t=100~\mathrm{yr}$. The gray dashed line denotes the reference Kolmogorov slope $P_\lambda(\lambda)\propto \lambda^{1/3}$.}
\label{fig:magnetic_spectrum_10-2}
\end{figure*}
As discussed in the previous section, when the background magnetic field is sufficiently weak, the NRH instability supplies magnetic fluctuations in a self-regulating manner. As a result, for each $B_0$, $E_{\max}$ converges to nearly the same value, almost independent of the initial fluctuation amplitude $\xi_{B,\mathrm{ini}}$. In contrast, for $B_0 \gtrsim 10^{-3}~\mathrm{G}$ this self-regulation breaks down, and $E_{\max}$ transitions to a regime determined by the initial conditions ($B_0,~\xi_{B,\mathrm{ini}}$). For example, $B_0=10^{-3}~\mathrm{G}$ corresponds to a boundary case, where the enhancement of $E_{\max}$ due to NRH instability is limited to a factor of 1.7 even for $\xi_{B,\mathrm{ini}}=0.01$ (Tab.~\ref{tab:Emax_compare_eta1e-4}). At $B_0=10^{-2}~\mathrm{G}$ the NRH instability ceases to operate entirely, and the final value of $E_{\max}$ is governed solely by both the background field strength $B_0$ and initial turbulence amplitude $\xi_{B,\mathrm{ini}}$.

The reduced efficiency of NRH instability arises from the significant decrease in the upstream CR current density $j_{\text{CR}}$. A smaller $j_{\text{CR}}$ increases the NRH growth timescale $t_{\text{NRH}}$ in Eq.~\eqref{eq:NRH_linear_timescale}, limiting the e-foldings achievable within the advection time $t_{\text{adv}}$. Fig.~\ref{fig:CR_current_density_magnetic_field_difference} shows that, for $(B_0,~\eta,~\xi_{B,\mathrm{ini}})=(10^{-2}~\mathrm{G},~10^{-4},~0.1)$, $j_{\text{CR}}$ at $t=100~\mathrm{yr}$ is more than two orders of magnitude smaller than in the $B_0=10^{-5}~\mathrm{G}$ case (Fig.~\ref{fig:CR_density_from_shock_10-5}), resulting in insufficient amplification as also summarized in Tab.~\ref{tab:NRH_timescale_compare}. The decline in $j_{\text{CR}}$ with increasing $B_0$ is explained by particle transport: a stronger magnetic field reduces the gyroradius for a CR of given energy, making it harder to leak into the upstream region. In the strong background magnetic field regime, NRH instability is therefore suppressed, and $E_{\max}$ is determined by the initial values of ($B_0,~\xi_{B,\mathrm{ini}}$).

Even for strong background magnetic fields ($B_0=10^{-2}~\mathrm{G}$), the acceleration efficiency agrees with the \nishiura{test-particle DSA} prediction (Eq.~\eqref{eq:order_estimation_maximum_energy_CR_Bell}) within a factor of 2. On the fluid side, however, clear damping of short-wavelength magnetic fluctuations appears, depending on the initial amplitude. As shown in Fig.~\ref{fig:magnetic_spectrum_10-2} (left), the case with $\xi_{B,\mathrm{ini}}=0.1$ exhibits strong damping for $\lambda \lesssim 10^{-2}~\mathrm{pc}$, whereas the case with $\xi_{B,\mathrm{ini}}=0.01$ (right) does not. This behavior is consistent with parametric instabilities, particularly stimulated Brillouin scattering, where a parent Alfv\'en wave decays into a backward Alfv\'en wave and a sound wave. The growth timescale of stimulated Brillouin scattering is expressed as follows \citep{1978ApJ...224.1013D,1978ApJ...219..700G,1993JGR....9819049J,2024PhRvE.110a5205I},
\begin{equation}
\begin{aligned}
t_\text{B} &= 2 \xi_{B}^{-\frac{1}{2}}(1+\theta) 
\sqrt{\frac{\theta}{1-\theta}} \frac{1}{k v_{\text{A}}} \\
&= 0.9~ \mathrm{yr}~
\left(\frac{\xi_{B}}{0.1}\right)^{-\frac{1}{2}}
\left(\frac{P_{\text {wind }}}{4.74 \times 10^{-7} ~\mathrm{dyn/cm}^2}\right)^{\frac{1}{4}} \\
&\times\left(\frac{B_0}{10^{-2}~ \mathrm{G}}\right)^{-\frac{3}{2}}
\left(\frac{\lambda}{10^{-2}~ \mathrm{pc}}\right)
\left(\frac{\rho_{\text {wind }}}{8.78 \times 10^{-24} ~\mathrm{g/cm}^3}\right)^{\frac{1}{2}},
\end{aligned}
\end{equation}
where $\theta$ is given by
\begin{equation}
\begin{aligned}
\theta &= \frac{c_\text{s}}{v_{\text{A}}}= 0.315
\left(\frac{P_{\text {wind }}}{4.74 \times 10^{-7}~ \mathrm{dyn/cm}^2}\right)^{\frac{1}{2}} \\
&\quad\times\left(\frac{B_0}{10^{-2}~ \mathrm{G}}\right)^{-1}
\left(\frac{\gamma}{5/3}\right)^{\frac{1}{2}}.
\end{aligned}
\end{equation}
The scaling $t_\text{B}\propto \xi_{B}^{-1/2}(k v_{\mathrm{A}})^{-1}$ implies $t_\text{B}\propto \delta B^{-1}$, so that stronger initial fluctuations ($\xi_{B,\mathrm{ini}}=0.1$) lead to faster damping at short wavelengths, while weaker fluctuations ($\xi_{B,\mathrm{ini}}=0.01$) suppress damping, consistent with the simulation results (Fig.~\ref{fig:magnetic_spectrum_10-2}).

In the right panel of Fig.~\ref{fig:magnetic_spectrum_10-2}, the short-wavelength magnetic energy spectra collapse to $P_\lambda(\lambda)\simeq \mathrm{const}$, which from Eq.~\eqref{eq:magnetic_power_spectrum_UFO} corresponds to $P_k(k)\propto k^{-2}$, consistent with both theoretical predictions and solar wind observations. Previous MHD simulations have shown that circularly polarized Alfv\'en waves with broadband spectra undergo time evolution such that, under magnetically dominated conditions ($c_{\mathrm{s}}\ll v_{\mathrm{A}}$), strong parametric decay into backward Alfv\'en waves and sound waves occurs, while the instability is suppressed under gas-pressure-dominated conditions ($c_{\mathrm{s}}\gg v_{\mathrm{A}}$) \citep{1992JGR....97.3113U,1996PhPl....3.4427M,2001NPGeo...8..159M}. Furthermore, \citet{2018JPlPh..84a9006C} derived, based on weak turbulence theory, that the scattered Alfv\'en-wave spectrum scales as $\omega^{-2}$. Observations also report that regions dominated by backward-propagating (sunward) scattered waves exhibit magnetic energy spectra with slopes close to $k^{-2}$ along the background field direction \citep{2008PhRvL.101q5005H,2009ApJ...698..986P,2011ApJ...733...76F}.
\begin{figure*}
\centering
\includegraphics[width=\textwidth]{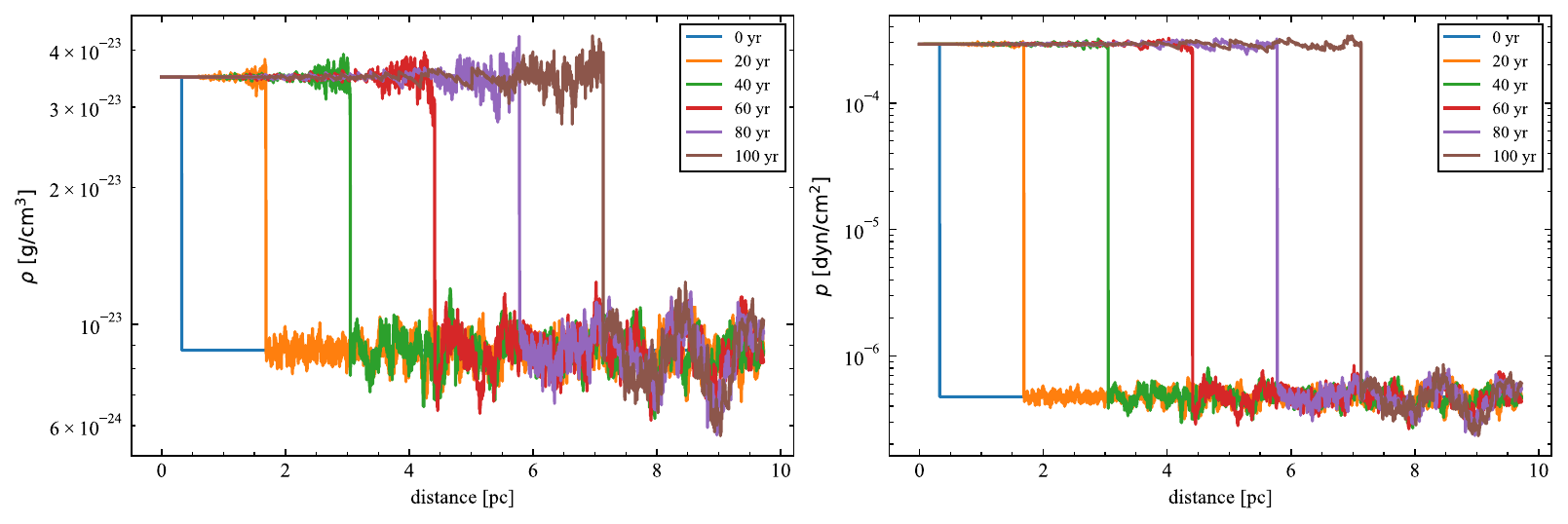}
\caption{\justifying 
Spatial distributions of fluid density (left) and fluid pressure (right) for the fiducial model with $B_0=10^{-2}~\mathrm{G}$ and $\eta=10^{-4}$, as defined in Tab.~\ref{tab:Model_Parameters_UFO}. The horizontal axis denotes the distance from the center in parsecs. Solid curves represent the profiles at $t=0~\mathrm{yr}$ (blue), $20~\mathrm{yr}$ (orange), $40~\mathrm{yr}$ (green), $60~\mathrm{yr}$ (red), $80~\mathrm{yr}$ (purple), and $100~\mathrm{yr}$ (brown).}
\label{fig:density_pressure_10^-2}
\end{figure*}

The decay of magnetic fluctuations is attributed to physical processes rather than numerical dissipation, supported by following three evidences. First, the shortest initial wavelength $\lambda_{\text{min}}\sim1.4\times10^{-3}~\text{pc}$ is resolved by 256 grid cells, and Eq.~\eqref{eq:condition_magnetic_fluctuation_divide} ensures that more than 95\% of the Alfv\'en wave amplitude remains. Second, the measured damping time is consistent with the timescale of stimulated Brillouin scattering. Third, as shown in Fig.~\ref{fig:density_pressure_10^-2}, density fluctuations appear in the density and pressure, and their amplitudes grow over time. This behavior can be interpreted as the growth of sound waves via stimulated Brillouin scattering.\footnote{A more rigorous identification of the decay mechanism requires Fourier analysis in both space and time, in order to verify whether peaks associated with stimulated Brillouin scattering appear on the dispersion relation of the sound wave.}

In the present simulations, the decay of magnetic fluctuations has only a limited impact on the efficiency of particle acceleration. This is because the Kolmogorov spectrum used as the initial condition places most of the energy at long wavelengths ($\sim$pc scale), which are largely unaffected by parametric instabilities. Even if the short-wavelength components decay, efficient particle acceleration can be maintained as long as the long-wavelength components persist.

However, this result depends on the idealized assumption that large-amplitude ($\xi_{B,\mathrm{ini}}\sim0.1$) long-wavelength fluctuations are always present near the UFO shock. In realistic environments, magnetic fluctuations generated near the black hole may undergo significant damping through parametric instabilities before propagating to pc scales, reducing the amplitude to $\xi_{B}\ll0.1$. Therefore, achieving particle acceleration up to $\sim10^{18}~\text{eV}$ in UFOs requires that sufficiently strong long-wavelength fluctuations survive to pc scales or that fresh turbulence is generated in situ.

\subsection{Variation of ISM Density}

This section examines the impact of ISM density on the efficiency of the NRH instability and the maximum CR acceleration energy (see the ISM low and ISM high models in Tab.~\ref{tab:Model_Parameters_UFO}). As shown in Tab.~\ref{tab:Emax_ISM_compare}, compared to the case with $n_{\text{ISM}}=10^2~\mathrm{cm}^{-3}$, the maximum acceleration energy $E_{\max}$ decreases by approximately one order of magnitude at $n_{\text{ISM}}=1~\mathrm{cm}^{-3}$, while it increases by a factor of a few at $n_{\text{ISM}}=10^4~\mathrm{cm}^{-3}$. The NRH instability becomes more efficient at higher ISM densities, leading to larger $E_{\max}$.

\begin{table}[htbp]
  \centering
  \renewcommand{\arraystretch}{1.5}
  \caption{Comparison of the maximum CR acceleration energy $E_{\max}$ at different ISM densities [eV]. 
  Column 1: ISM density $n_{\mathrm{ISM}}$ [cm$^{-3}$]. 
  Column 2: without NRH (No NRH). 
  Column 3: with NRH for $\eta=10^{-4}$. 
  Column 4: amplification factor defined as the ratio between the NRH and No NRH cases.}
  \label{tab:Emax_ISM_compare}
\begin{tabular}{lccc}
  \toprule
  $n_{\mathrm{ISM}}$ & No NRH & $\eta=10^{-4}$ & Amplification \\
  \midrule
  $1$    & $1.1\times10^{14}~\mathrm{eV}$ & $2.2\times10^{14}~\mathrm{eV}$ & $2.0$ \\
  $10^2$ & $7.9\times10^{14}~\mathrm{eV}$ & $2.2\times10^{15}~\mathrm{eV}$ & $2.8$ \\
  $10^4$ & $7.9\times10^{14}~\mathrm{eV}$ & $3.6\times10^{15}~\mathrm{eV}$ & $4.6$ \\
  \bottomrule
\end{tabular}
\end{table}
\begin{figure}
\RaggedRight
\includegraphics[width=\columnwidth]{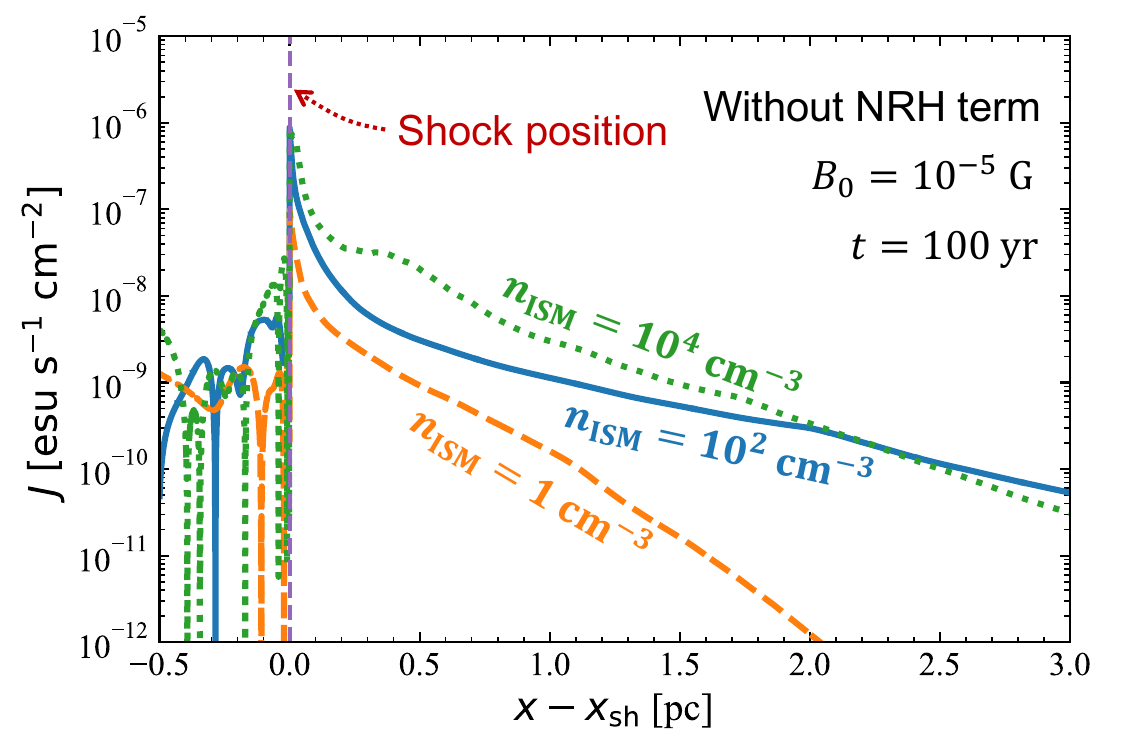}
\caption{\justifying CR current profiles escaping upstream of the shock for different ISM densities. The parameters are $(B_0,\,\xi_{B,\mathrm{ini}},\,t)=(10^{-5}~\mathrm{G},\,0.1,\,100~\mathrm{yr})$. 
The horizontal axis represents the distance from the shock (in pc), and the vertical axis shows the CR current (in $\mathrm{esu}~\mathrm{s}^{-1}~\mathrm{cm}^{-2}$). 
The orange dashed, blue solid, and green dotted lines correspond to ISM densities of $1$, $10^{2}$, and $10^{4}~\mathrm{cm^{-3}}$, respectively. 
The purple dashed line indicates the shock position.
}
\label{CR_current_different_ISM}
\end{figure}

The maximum CR energy depends strongly on the reverse shock velocity (see Eq.~\eqref{eq:order_estimation_maximum_energy_CR_Bell}). As the ISM density increases, the shock velocity in the wind rest frame also increases. In the extreme limit $n_{\mathrm{ISM}}\rightarrow\infty$, corresponding to complete reflection of the reverse shock by the ISM, the upstream velocity in the shock frame becomes $v_{\text{u}}=v_{\text{wind}}+v_{\text{sh}}$ and the downstream velocity is $v_{\text{d}}=v_{\text{sh}}$. With $v_{\text{wind}}=0.2c$ and a compression ratio of 4, one obtains $v_{\text{u}}=4v_{\text{d}}$. Thus, the shock velocity in the lab frame is $v^{\prime}_{\text{sh}}=1/3v_{\text{wind}}$, and in the wind rest frame it becomes $v_{\text{sh}}=4/3v_{\text{wind}}=8\times10^9~\mathrm{cm~s}^{-1}$. Therefore, even as ISM density increases indefinitely, the shock velocity saturates at this value. Tab.~\ref{tab:shocked_properties_compare} shows that as the ISM density increases, the rise in shock velocity becomes smaller, with little change between $n_{\text{ISM}}=10^2$ and $10^4~\mathrm{cm}^{-3}$. Consequently, $E_{\max}$ also shows a saturation trend, consistent with the scaling $E_{\max}\propto v_{\text{sh}}^2$ in Eq.~\eqref{eq:order_estimation_maximum_energy_CR_Bell}.

\begin{table}[htbp]
  \centering
  \renewcommand{\arraystretch}{1.5}
  \caption{Fluid quantities in the reverse-shocked region (wind rest frame) at different ISM densities. 
  Column 1: ISM density $n_{\text{ISM}}$ [cm$^{-3}$]. 
  Column 2: shocked density $\rho_{\text{shocked}}$ [g cm$^{-3}$]. 
  Column 3: shocked pressure $P_{\text{shocked}}$ [dyn cm$^{-2}$]. 
  Column 4: shock velocity $v_{\text{sh}}$ [cm s$^{-1}$].}
  \label{tab:shocked_properties_compare}
  \begin{tabular}{lccc}
    \toprule
    $n_{\text{ISM}}$ & Density & Pressure & $v_{\text{sh}}$ \\
    \midrule
    $1$     & $3.4\times10^{-23}$  & $4.6\times10^{-5}$  & $2.0\times10^{9}$  \\
    $10^2$  & $3.5\times10^{-23}$  & $2.9\times10^{-5}$  & $5.0\times10^{9}$  \\
    $10^4$  & $3.5\times10^{-23}$  & $4.1\times10^{-4}$  & $5.9\times10^{9}$  \\
    \bottomrule
  \end{tabular}
\end{table}

In our simulations, increasing $n_{\mathrm{ISM}}$ leads to higher $v_{\text{sh}}$, as shown in Tab.~\ref{tab:shocked_properties_compare}, which enhances the efficiency of the NRH instability. 
A larger $v_{\text{sh}}$ reduces the CR acceleration time in Eq.~\eqref{eq:acceleration_time_UFO}, since $t_{\text{acc}}\propto v_{\text{sh}}^{-2}$ when $B_0$, $\xi_{B,\mathrm{ini}}$, and $E_{\mathrm{CR}}$ are fixed. 
Consequently, $t_{\text{acc}}$ becomes shorter, allowing high-energy CRs to be produced more rapidly. 
These CRs escape further upstream, increasing the CR current density $j_{\text{CR}}$ ahead of the shock. 
Figure~\ref{CR_current_different_ISM} illustrates the spatial distribution of the upstream CR current at $t=100~\mathrm{yr}$ for different ISM densities. 
It clearly shows that a higher ISM density results in a larger CR current escaping into the upstream region. 
As $j_{\text{CR}}$ increases, the NRH instability growth time $t_{\text{NRH}}$ in Eq.~\eqref{eq:NRH_linear_timescale} decreases, which results in more e-foldings $t_{\text{adv}}/t_{\text{NRH}}$ and stronger magnetic amplification.

\subsection{Effect of p$\gamma$ Cooling}
\label{subsec:pgamma_cooling_UFO}
\begin{figure*}
\centering
\includegraphics[width=\textwidth]{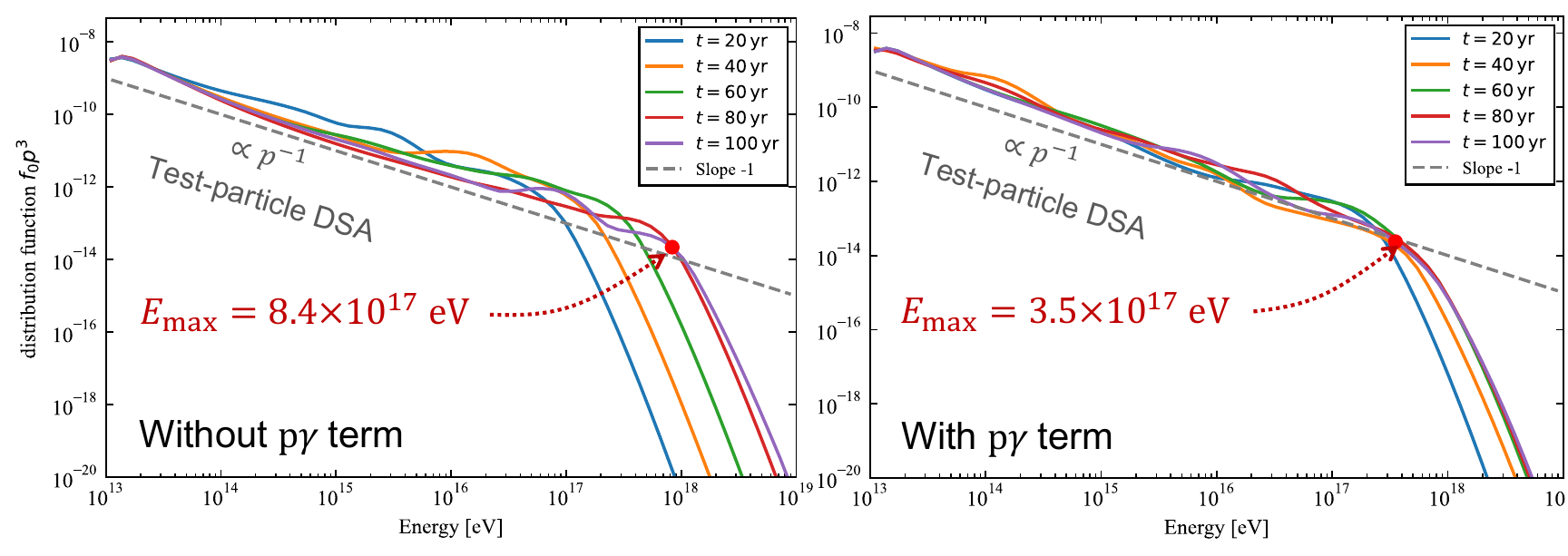}
\caption{\justifying 
Energy spectrum of the isotropic CR distribution function in the $\mathrm{p}\gamma$ model (see Tab.~\ref{tab:Model_Parameters_UFO}) for $B_0=10^{-2}~\mathrm{G}$. 
Left: without $\mathrm{p}\gamma$ cooling. 
Right: with $\mathrm{p}\gamma$ cooling included. 
In both cases, the NRH term is taken into account. 
The blue, orange, green, red, and purple solid lines denote the spectra at $t=20$, $40$, $60$, $80$, and $100~\mathrm{yr}$, respectively, measured from the start of the simulation. 
The gray dashed line indicates the slope expected from the analytic DSA solution. 
The red dots mark the maximum CR energy, defined as the momentum where the spectral index of $f_0p^3$ falls below $-2.1$.}
\label{fig:CR_spectrum_pgamma}
\end{figure*}
\begin{table}[htbp]
  \centering
  \renewcommand{\arraystretch}{1.5}
  \caption{Maximum CR acceleration energy $E_{\max}$ [$\mathrm{eV}$] in the $\mathrm{p}\gamma$ model with NRH term (see Tab. \ref{tab:Model_Parameters_UFO}). 
  Column 1: background magnetic field $B_0$. 
  Column 2: without $\mathrm{p}\gamma$ cooling (No $\mathrm{p}\gamma$). 
  Column 3: with $\mathrm{p}\gamma$ cooling (With $\mathrm{p}\gamma$).}
  \label{tab:Maximum_Energy_pgamma}
\begin{tabular}{lcc}
  \toprule
  $B_0$ [G] & No $\mathrm{p}\gamma$ & With $\mathrm{p}\gamma$ \\
  \midrule
  $10^{-4}$ & $9.8\times10^{15}~\mathrm{eV}$ & $9.8\times10^{15}~\mathrm{eV}$ \\
  $10^{-3}$ & $9.5\times10^{16}~\mathrm{eV}$ & $7.4\times10^{16}~\mathrm{eV}$ \\
  $10^{-2}$ & $8.4\times10^{17}~\mathrm{eV}$ & $3.5\times10^{17}~\mathrm{eV}$ \\
  \bottomrule
\end{tabular}
\end{table}
In this subsection, we examine the impact of $\mathrm{p}\gamma$ cooling on the maximum CR acceleration energy with NRH instability. 
As shown in Tab.~\ref{tab:Maximum_Energy_pgamma}, $\mathrm{p}\gamma$ cooling becomes effective when the background magnetic field increases to $B_0 \sim 10^{-3}~\mathrm{G}$, corresponding to $E_{\max}\sim10^{17}~\mathrm{eV}$. 
The suppression is most prominent for $B_0=10^{-2}~\mathrm{G}$. 
Fig.~\ref{fig:CR_spectrum_pgamma} illustrates this behavior: without $\mathrm{p}\gamma$ cooling, CRs reach $\sim\mathrm{EeV}$ ($10^{18}~\mathrm{eV}$), whereas with cooling, the maximum energy is reduced to the sub-EeV regime. 
In this case, $E_{\max}$ saturates at a nearly constant value after $t \sim 60~\mathrm{yr}$.

This behavior can be interpreted from the comparison of timescales in Fig.~\ref{fig:pgamma_cooling_timescale}. 
In standard DSA theory, the maximum energy is determined where the acceleration timescale $t_{\mathrm{acc}}$ (Eq.~\eqref{eq:acceleration_time_UFO}) equals the simulation time $t_{\mathrm{final}}$. 
In Fig.~\ref{fig:pgamma_cooling_timescale}, this condition corresponds to the intersection of the red dot-dashed line ($t_{\mathrm{acc}}$) and the orange dashed line ($t_{\mathrm{final}}$). 
When the $\mathrm{p}\gamma$ cooling timescale $t_{\mathrm{p}\gamma}$ becomes shorter than $t_{\mathrm{final}}$, however, $E_{\max}$ is instead determined by the intersection of $t_{\mathrm{acc}}$ with $t_{\mathrm{p}\gamma}$ (blue solid line). 
For the $B_0=10^{-2}~\mathrm{G}$ model, the maximum energy estimated from $t_{\mathrm{acc}}=t_{\mathrm{final}}$ is $\sim1.8\times10^{18}~\mathrm{eV}$, while that from $t_{\mathrm{acc}}=t_{\mathrm{p}\gamma}$ is $\sim7.1\times10^{17}~\mathrm{eV}$. 
Thus, $\mathrm{p}\gamma$ cooling reduces $E_{\max}$ by a factor of $\sim0.4$, in good agreement with the numerical results in Fig.~\ref{fig:CR_spectrum_pgamma}.

The condition for $\mathrm{p}\gamma$ cooling to impose a limit on $E_{\max}$ can be expressed in terms of a critical magnetic field strength $B_{\mathrm{p}\gamma}$. 
This corresponds to the point where $t_{\mathrm{acc}}$, $t_{\mathrm{final}}$, and $t_{\mathrm{p}\gamma}$ intersect. 
Using $t_{\mathrm{final}}=100~\mathrm{yr}$ and the intersection energy $E_{\mathrm{CR}}=1.3\times10^{16}~\mathrm{eV}$ between $t_{\mathrm{final}}$ and $t_{\mathrm{p}\gamma}$, we estimate
\begin{equation}
\begin{aligned}
    B_{\mathrm{p}\gamma} &> 7.1\times10^{-5}~ \mathrm{G} \\
    & \times \left(\frac{t_{\mathrm{final}}}{100~  \mathrm{yr}}\right)^{-1}
    \left(\frac{v_{\mathrm{sh}}}{5.0\times10^9~\text{cm/s}}\right)^{-2}\left(\frac{\xi_{B}}{0.1}\right)^{-1}.
\end{aligned}
\end{equation}

\nishiura{Note that in Figs.~\ref{fig:CR_spectrum_fiducial} (the run without the NRH term) and \ref{fig:CR_spectrum_pgamma}, the CR spectrum approximately follows the test-particle DSA slope $f(p)\propto p^{-4}$, with small wiggles around this power law. We think there two main reasons for the wiggling. First, the amplitude of the magnetic fluctuations $\delta B$ varies with position and time, so the diffusion coefficient in Eq.~\eqref{eq:diffusion_coefficient_pitch_angle} becomes space- and time-dependent. This variation changes the number of CRs that remain confined in, or escape from, the acceleration region and produces small deviations from a test-particle spectrum. Second, in the strong $B_0=10^{-2}~\mathrm{G}$ models in Tab.~\ref{tab:Model_Parameters_UFO}, parametric decay instabilities generate large density and pressure fluctuations (see Fig.~\ref{fig:density_pressure_10^-2}). According to Eq.~\eqref{eq:diffusion_length_UFO_bunkai}, CRs with different momenta have different diffusion lengths and therefore sample different parts of this inhomogeneous shock structure, experiencing slightly different effective compression ratios. By contrast for weak $B_0$ case, as shown in Fig.~\ref{fig:CR_spectrum_fiducial} (the run with the NRH term), small-scale magnetic fluctuations generated by the NRH instability dominate over the initial turbulence, so the scattering environment is more homogeneous and the spectrum tends to be smoother.}

\section{Conclusion}

In this work, we evaluated the growth of NRH instability–driven magnetic turbulence and the maximum CR acceleration energy $E_{\mathrm{max}}$ in reverse shocks of AGN UFOs. Using a numerical framework that self-consistently couples the CR diffusion--convection equation with nonlinear MHD evolution, we examined the dependence on $B_0$, initial strength of magnetic turbulence $\xi_{B,\mathrm{ini}}$ in Eq. \eqref{eq:definition_magnetic_fluctuation_strength}, and injection rate $\eta$, while including the effects of NRH instability and $\mathrm{p}\gamma$ cooling.

Previous PIC simulations at very low CR maximum energies demonstrated strong magnetic field amplification due to the NRH instability that saturates in nonlinear stage \citep{2014ApJ...783...91C,2014ApJ...794...46C}. By contrast, \citet{2021ApJ...922....7I,2024ApJ...965..113I} showed only moderate amplification in simulations of SNRs consistent with observations. In their simulations, the maximum CR energies were~$1~\mathrm{PeV}$ or less. Our study demonstrates that when the maximum energy is even higher ($\lesssim1 ~\mathrm{EeV}$), magnetic field amplification due to NRH instability becomes much weaker. \nishiura{Previous PIC simulations have shown that at Alfvén Mach numbers \(\mathcal{M}_{\mathrm{A}}\sim10^{2}\), the magnetic field can be amplified up to \(\xi_{B}\sim60\) \citep{2014ApJ...794...46C}. By contrast, in our setup the NRH instability is largely ineffective for \(\mathcal{M}_{\mathrm{A}}<10^{2}\) (see Tab. \ref{tab:Emax_compare_eta1e-4}).} Since the CR current originates from particles escaping upstream near $E_{\max}$, it is natural that the current diminishes with increasing maximum energy, thereby suppressing NRH instability growth. Many earlier works did not explicitly examine the dependence of magnetic amplification on maximum energy, but our results suggest that realistic theoretical models must describe NRH amplification as a function of $E_{\max}$.

To accelerate CRs up to $\sim10^{18}~\mathrm{eV}$ in UFOs, the following conditions must be simultaneously satisfied, which is not as easy as previously thought:
\begin{enumerate}[label=(\roman*)]
 \item Near the reverse shock (within $\sim1~\mathrm{pc}$), both the background magnetic field $B_0$ and turbulent amplitude $\xi_{B,\mathrm{ini}}$ must be sufficiently large, specifically $B_0\geq10^{-2}~\mathrm{G}$ and $\xi_{B,\mathrm{ini}}\geq0.1$. Under these conditions, the NRH instability is ineffective, and the local parameters of the acceleration region determine $E_{\max}$. In contrast, when the background magnetic field is weak ($B_0<10^{-4}~\mathrm{G}$), the NRH instability operates efficiently, but the acceleration efficiency is insufficient to reach the EeV range.
 \item The magnetic fluctuation spectrum must be dominated by long wavelengths (Kolmogorov-type). If short wavelengths dominate initially, parametric instabilities such as stimulated Brillouin scattering cause their rapid decay, reducing the acceleration efficiency (Fig.~\ref{fig:magnetic_spectrum_10-2}).
 \item The $\mathrm{p}\gamma$ cooling must remain inefficient (i.e., the AGN photon field must be weak). 
\citet{2023MNRAS.526..181P} showed that EeV-scale acceleration is possible even with $\mathrm{p}\gamma$ cooling if the magnetic energy fraction is high ($\epsilon_B \simeq 0.05$). 
In our simulations, the downstream magnetic energy fraction eventually reaches $\epsilon_B^{\mathrm{down}}\simeq0.04$ for the initial condition $(B_0,\,\xi_{B,\mathrm{ini}})=(10^{-2}~\mathrm{G},\,0.1)$ (see Tab.~\ref{tab:epsB_up_down_summary}), similar to their assumed acceleration conditions. 
However, the maximum CR energy remains sub-EeV (Fig.~\ref{fig:CR_spectrum_pgamma}). 
Thus, achieving EeV energies likely requires a photon field weaker than that in \citet{2023MNRAS.526..181P}. 
The difference is only a factor of a few and is therefore not so serious.

\end{enumerate}
Whether these conditions are realized in UFO environments requires future observational confirmation.

In the weak magnetic field regime ($B_0\leq10^{-4}~\mathrm{G}$), the NRH instability self-consistently amplifies magnetic fluctuation $\xi_{B}$ regardless of $\xi_{B,\mathrm{ini}}$, and $E_{\max}$ is automatically determined. For $\eta=10^{-4}$, $E_{\max}$ reaches $\sim10^{16}~\mathrm{eV}$ at $B_0=10^{-4}~\mathrm{G}$ and $\sim10^{15}~\mathrm{eV}$ at $B_0=10^{-5}~\mathrm{G}$. At a higher injection rate $\eta=10^{-3}$, Tab.~\ref{tab:Maximum_Energy_Fiducial} shows that $E_{\max}$ can increase further by a factor of a few.

A transition occurs at $B_0\gtrsim10^{-3}~\mathrm{G}$, where the linear NRH instability growth time $t_{\mathrm{NRH}}$ (Eq.~\eqref{eq:NRH_linear_timescale}) tends to exceed the upstream advection time $t_{\mathrm{adv}}$ (Eq.~\eqref{eq:definition_advection_time_Bell}), so that the NRH instability is ineffective. In this regime, $E_{\max}$ is controlled by the initial conditions ($B_0,\ \xi_{B,\mathrm{ini}}$) and by $\mathrm{p}\gamma$ cooling (Tabs.~\ref{tab:Maximum_Energy_Fiducial}, \ref{tab:Maximum_Energy_pgamma}). The key reason for $t_{\mathrm{adv}}/t_{\mathrm{NRH}}<1$ is simply due to the suppression of the escaping CR current $j_{\mathrm{CR}}$ at larger $B_0$. A stronger magnetic field decreases the gyroradius, making it harder for CRs of a given energy to escape far upstream.

At $B_0\sim10^{-2}~\mathrm{G}$, acceleration up to sub--$\mathrm{EeV}$ is possible if the initial magnetic turbulence $\xi_{B,\mathrm{ini}}$ is long-wavelength dominated. However, if, in realistic UFO environments, most of the magnetic fluctuation energy may resides at scales $\lambda\lesssim10^{-2}~\mathrm{pc}$, the acceleration efficiency would be strongly reduced. Such short-wavelength fluctuations decay through parametric instabilities, including stimulated Brillouin scattering, and $E_{\max}$ would then fall below $10^{17}~\mathrm{eV}$.

Recent XRISM results indicate mass outflow rates of $\sim100~M_{\odot}\,\mathrm{yr}^{-1}$ \citep{2025Natur.641.1132X}, and both theory and observations suggest that UFOs are clumpy \citep{2013PASJ...65...88T,2025Natur.641.1132X}. Future work must assess how the large mass outflow rates and clumpy structure of UFOs affect efficiency of NRH instability and CR acceleration. Higher mass outflow rates may increase the number of particles contributing to the CR current, thereby enhancing NRH instability. If UFOs are generally clumpy and inhomogeneous, shock geometry would become highly non-uniform, and the acceleration efficiency could differ substantially from the uniform-wind case (e.g., \citet{2012ApJ...744...71I} for SNR case). Addressing these issues will require combined theoretical, numerical, and observational efforts.

\begin{acknowledgments}
\nishiura{We are grateful to the anonymous referee for constructive comments and suggestions that improved the manuscript.} We also thank \nishiura{Yutaka Ohira, Shoma F. Kamijima}, Misaki Mizumoto, Susumu Inoue, Kohta Murase, Kunihito Ioka, Wataru Ishizaki and Nobuyuki Sakai for fruitful discussions that greatly advanced this work. This work is supported by JST SPRING Grant No. JPMJSP2110 (RN), JSPS KAKENHI, Grant No. 25KJ1562 (RN), and Grants-in-aid from the Ministry of Education, Culture, Sports, Science, and Technology (MEXT) of Japan No. 20H01944 and 23H01211 (TI). This work was performed using the Yukawa-21 supercomputer at the Yukawa Institute for Theoretical Physics, Kyoto University, and the supercomputing facilities of the Center for Computational Astrophysics, National Astronomical Observatory of Japan, including the XC-50 and XD-2000 systems.

\end{acknowledgments}

\bibliography{sample701}{}
\bibliographystyle{aasjournalv7}

\end{document}